\documentclass[journal]{IEEEtran}
\usepackage{amsmath,amsfonts}
\usepackage{algorithmic}
\usepackage{algorithm}
\usepackage{array}
\usepackage[caption=false]{subfig}
\usepackage{textcomp}
\usepackage{stfloats}
\usepackage{url}
\usepackage{verbatim}
\usepackage{graphicx}
\usepackage{cite}
\usepackage{color}
\usepackage{amssymb}
\usepackage{multirow}
\usepackage{multicol}

\newtheorem{theorem}{Theorem}

\newtheorem{lemma}{Lemma}

\newtheorem{remark}{Remark}

\DeclareMathOperator{\sinc}{sinc}

\begin{document}

\title{On the Time-Frequency Localization Characteristics of the Delay-Doppler Plane Orthogonal Pulse}

\author{Akram Shafie, \IEEEmembership{Member, IEEE,} Jinhong Yuan, \IEEEmembership{Fellow, IEEE,} Nan Yang, \IEEEmembership{Senior Member, IEEE,}\\and Hai Lin, \IEEEmembership{Senior Member, IEEE}
\thanks{The work of A. Shafie and J. Yuan was supported in part by the Australian Research Council (ARC) Discovery Project under Grant DP220103596, and in part by the ARC Linkage Project under Grant LP200301482. The work of H. Lin was supported by the Japan Society for the Promotion of Science (JSPS) Grants-in-Aid for Scientific Research (KAKENHI) under Grant 22H01491.}
\thanks{A. Shafie and J. Yuan are with the School of Electrical Engineering and Telecommunications, The University of New South Wales, Sydney, NSW, 2052, Australia (e-mail: akram.shafie@unsw.edu.au, j.yuan@unsw.edu.au).}
\thanks{N. Yang is with the School of Engineering, The Australian National University, Canberra, ACT 2601, Australia (e-mail: nan.yang@anu.edu.au).}
\thanks{H. Lin is with the Graduate School of Engineering, Osaka Metropolitan University, Sakai, Osaka, 599-8531, Japan (e-mail: hai.lin@ieee.org).}
\thanks{A preliminary version of this work was presented \textcolor{black}{ in 2024 IEEE Global Communications Conference (Globecom)}~\cite{AS_2024_GC_ODDMTFLoc}.}}


\maketitle

\begin{abstract}
In this work, we study the time-frequency (TF) localization characteristics of the prototype pulse of orthogonal delay-Doppler (DD) division multiplexing modulation, namely, the DD plane orthogonal pulse (DDOP). The TF localization characteristics examine how concentrated or spread out the energy of a pulse is in the joint TF domain, the time domain (TD), and the frequency domain (FD). We first derive the TF localization metrics of the DDOP, including its TF area, its time and frequency dispersions, and its direction parameter. Based on these results, we demonstrate that the DDOP exhibits a high energy spread in the TD, FD, and the joint TF domain, while adhering to the Heisenberg uncertainty principle. Thereafter, we discuss the potential advantages brought by the energy spread of the DDOP, especially with regard to harnessing both time and frequency diversities and enabling fine-resolution sensing. Subsequently, we examine the relationships between the time and frequency dispersions of the DDOP and those of the \textcolor{black}{envelope functions of DDOP’s TD and FD representations, paving the way for simplified determination of the TF localization metrics for more generalized variants of the DDOP and the pulses used in other DD domain modulation schemes.} 
Finally, using numerical results, we validate our analysis and find further insights. 
\end{abstract}


\section{Introduction}


\IEEEPARstart{O}{rthogonal} delay-Doppler (DD) division multiplexing (ODDM) modulation has recently been proposed as a promising multi-carrier modulation scheme for achieving reliable communications in high-mobility scenarios~\cite{2022_TWC_JH_ODDM}. The ODDM modulation couples the modulated signal with the DD domain representation of the doubly-selective channel,  which possesses inherent stability and sparsity characteristics~\cite{2021_WCM_JH_OTFS,2023_IoTJ_OTFSRISOutperformOFDM}. This efficient coupling empowers ODDM modulation to effectively minimize interference and exploit both time and frequency diversities, enabling reliable communications in high-mobility scenarios with low pilot overhead and low receiver processing complexity. 

At the core of ODDM modulation lies its prototype pulse, called the DD plane orthogonal pulse (DDOP). As outlined in~\cite{2022_ICC_JH_ODDM,2022_TWC_JH_ODDM}, the DDOP is constructed by concatenating multiple sub-pulses (see Fig. \ref{Fig:ut}).
The sub-pulse can be any square root-Nyquist (SRN) pulse that is parameterized by the zero inter-symbol-interference (ISI) interval which is equal to the fine time (delay) resolution of ODDM modulation. 
\textcolor{black}{We highlight that ODDM modulation and the widely explored DD domain modulation scheme, known as orthogonal time-frequency (TF) space  (OTFS) modulation, has the same digital sequence before pulse shaping~\cite{2017_WCNC_OTFS_Haddani,AS_2024_TCOM_OTFSOFDMCoexistence}. Despite this, thanks to the improved out-of-band emission (OOBE) performance and orthogonality characteristics of DDOP compared to the rectangular pulse used in OTFS modulation, the ODDM modulation exhibits superior bit error performance compared to OTFS modulation~\cite{2022_TWC_JH_ODDM}.}

Due to its benefits,~\cite{2022_ICC_JH_ODDM,2022_TWC_JH_ODDM,2022_GC_JH_ODDMPulse,2023_ODDM_TCOM} investigated the properties of the DDOP in various contexts. In particular,~\cite{2022_ICC_JH_ODDM,2022_TWC_JH_ODDM} proved the orthogonality of the DDOP with respect to (w.r.t.) the fine time (delay) and fine frequency (Doppler) resolutions of ODDM modulation. Moreover, by introducing a local or sufficient orthogonality for the DDOP corresponding to the Weyl-Heisenberg subset,~\cite{2022_GC_JH_ODDMPulse,2023_ODDM_TCOM} extended the orthogonality proof in~\cite{2022_ICC_JH_ODDM,2022_TWC_JH_ODDM} to the scenario where
the duration constraint of sub-pulses is released.  \textcolor{black}{Furthermore,~\cite{2023_ODDM_TCOM} analyzed the bandwidth efficiency of the DDOP and compared it with those of the pulses used in time division multiplexing (TDM) and frequency division multiplexing (FDM) schemes.} Additionally,~\cite{2022_GC_JH_ODDMPulse,2023_ODDM_TCOM} derived the frequency response of the DDOP and showed that any given DDOP can be generated by using a rectangular time domain (TD) window and a filter that has the impulse response of the sub-pulse. Despite these preliminary endeavors, the literature lacks a comprehensive discussion on the TF localization characteristics of the DDOP~\cite{2014_ComSurvTut_MCCommunication,1997_Hass_TFLocalization,Book_TheoryofCommunications_Gabor1946}.
This motivates us to explore the TF localization characteristics of the DDOP in detail, thereby filling a significant gap in the existing literature.

The TF localization characteristics examine how concentrated or spread out the energy of a pulse is in the joint TF domain, the TD, and the frequency domain (FD)~\cite{2003_TFLocalizationMultiCarrier}. 
In the literature, several classical metrics are used to examine the TF localization characteristics of a pulse. \textcolor{black}{Particularly, the metrics including the TF area (TFA) $\Delta A$, time dispersion $\Delta T$, frequency dispersion $\Delta F$, and direction parameter $\kappa$ are widely used,}\footnote{It is worth noting that in certain studies, the TFA, time dispersion, and frequency dispersion are referred to as dispersion product, pulse's effective duration, and pulse's effective bandwidth, respectively~\cite{2017_JSAC_FBMC_forLocalization,2014_ComSurvTut_MCCommunication,1997_Hass_TFLocalization,Book_TheoryofCommunications_Gabor1946}.}    serving as quantifiable measures of the energy spread of a pulse in the joint TF domain, the TD, the FD, and the TD relative to the FD, respectively~\cite{2023_ODDM_TCOM,2011_SignProcMag_FilterBankMultiCarrier,2017_JSAC_FBMC_forLocalization,2014_ComSurvTut_MCCommunication}.
As a consequence of the Heisenberg uncertainty principle, \textcolor{black}{TFA of any practically realizable pulse} adheres to a lower bound known as the Gabor limit~\cite{Book_TheoryofCommunications_Gabor1946}. Typically, a pulse is considered to possess \textcolor{black}{minimal energy spread in the joint TF domain} (or be well-localized in the joint TF domain) 
if its TFA approaches the Gabor limit~\cite{2017_JSAC_FBMC_forLocalization,2014_ComSurvTut_MCCommunication}.
Moreover, a pulse is considered to exhibit a higher energy spread in the TD (or FD) than another pulse when its time (or frequency) dispersion is larger compared to the other pulse. \textcolor{black}{Furthermore, a larger direction parameter indicates that the pulse is stretched more along the TD compared to the FD~\cite{2017_JSAC_FBMC_forLocalization,2015_TFLocalizationforBOTDR,2014_ComSurvTut_MCCommunication}.}

\color{black}
Examining the TF localization metrics of pulses and performing subsequent analyses offers several benefits. First and foremost, this examination provides us with a comprehensive and profound understanding of the energy spread of pulses in both TD and FD, which is crucial for assessing their potential in exploiting channel diversity and achieving accurate channel estimation for communication and sensing applications.
We clarify that pulses with either wide or narrow energy spread have distinct advantages and disadvantages for communication and sensing applications. Specifically, pulses with wide energy spread in the TD (FD) enhance the ability to harness the time (frequency) diversity of the channel, while those with narrow energy spread in the TD (FD) enable the possibility of more accurate estimation of delay (Doppler) response of wireless channels. Understanding these benefits necessitates a thorough examination of the TF localization metrics of pulses.
\color{black}

\color{black}
Another benefit of examining the TF localization characteristics is to allow us verifying whether the TFA of pulses $\Delta A$ adheres to the theoretical limits imposed by the Heisenberg uncertainty principle, thereby confirming their practical existence~\cite{2003_TFLocalizationMultiCarrier}. This is particularly relevant in the realm of DD domain modulation schemes (such as ODDM, OTFS, etc.), where it is commonly believed that pulses well-suited for such schemes may not exist in practice as the delay and Doppler resolutions of DD domain modulation schemes are extremely small~\cite{2017_WCNC_OTFS_Haddani,2022_GC_JH_ODDMPulse}. 

\color{black}

Several previous studies have explored the TF localization characteristics of various well-established pulses~\cite{1997__MCSurveyRef85,2003_TCOM_MCSurveyRef18,2007_Thesis_MCSurveyRef25,2011_SignProcMag_FilterBankMultiCarrier,2017_JSAC_FBMC_forLocalization,2010_TFLocalizationforWavelet,2015_TFLocalizationforBOTDR}. These include the examination of TF localization characteristics for orthogonal frequency division multiplexing (OFDM) pulses in~\cite{1997__MCSurveyRef85,2003_TCOM_MCSurveyRef18,2007_Thesis_MCSurveyRef25}, filter bank multicarrier waveforms in~\cite{2011_SignProcMag_FilterBankMultiCarrier,2017_JSAC_FBMC_forLocalization}, biorthogonal wavelets in~\cite{2010_TFLocalizationforWavelet}, and pulses in Brillouin optical TD reflectometry systems in~\cite{2015_TFLocalizationforBOTDR}. While recognizing their importance, we note that none of these studies have delved into exploring the TF localization characteristics of the DDOP, primarily due to the recent discovery of DDOP.

In this work, we conduct a comprehensive analysis of the TF localization characteristics of the DDOP. The key contributions of this work are summarized as follows:
\begin{itemize}
\item
We first derive the TF localization metrics of the DDOP, including its TFA $\Delta A$, time dispersion $\Delta T$, and frequency dispersion $\Delta F$, and the direction parameter $\kappa$. In particular, we derive these metrics when its sub-pulse is the truncated version of either the root-raised-cosine (RRC) pulse or the better-than RRC (BTRRC) pulse.
\item
By analyzing the derived TF localization metrics, 
we provide valuable insights regarding the energy spread of the DDOP. 
 In particular, we show  that
\begin{itemize}
  \item Energy spread of the DDOP is high in both TD and FD, and these lead to its energy spread in the joint TF domain to be high. These characteristics are different from those of the pulses in TDM and FDM schemes, which exhibit wide energy spread in one domain (either TD or FD), and low energy spread in the other,  leading to relatively low energy spread in the joint TF domain;
  \item The DDOP does not violate the Heisenberg uncertainty principle, thereby confirming the practical existence of pulses well-suited for DD domain modulation schemes;
  \item Due to the scattered nature of its energy distribution in the TD, FD, and joint TF domain, DDOP behaves locally like a pulse with narrow energy spread in the TD, the FD, and the joint TF domain, even though its energy spread is globally high in these domains;
  \item the DDOP exhibits potential advantages due to its energy spread, especially with regards to harnessing both time and frequency diversities, as well as enabling sensing with fine resolutions.
\end{itemize}
\item
\textcolor{black}{By studying the envelope functions of DDOP's TD and FD
representations,} we arrive at the following insights: 
\begin{itemize}
  \item The time dispersion $\Delta T$ of the DDOP can be determined based on $\Delta T$ of the envelope function of DDOP’s TD representation;
  \item The frequency dispersion $\Delta F$ of the DDOP can be determined based on $\Delta F$ of the envelope function of DDOP’s FD representation.
\end{itemize}
Based on these findings, we determine the TF localization metrics of the recently proposed generalized design of the DDOP and effective pulses in OTFS. 
\item
Based on numerical results, we verify our analysis by comparing it with simulation results. Also, we show that the energy spread in the joint TF domain and TD of the generalized design of the DDOP exhibit a step-wise increase as the duration of sub-pulses increases. Furthermore, we show that the DDOP can achieve a lower energy spread in the joint TF domain and FD, if the RRC pulse, as opposed to the BTRRC pulse, is utilized as its sub-pulse.
\end{itemize}

The rest of the paper is organized as follows: Section II covers the preliminaries of ODDM modulation, DDOP, and the TF localization characteristics of a pulse. Section III presents the derivation of TF localization metrics for DDOP. Section IV delves into discussion, potential benefits, and insights related to our TF localization characteristics. Sections V and VI discuss the TF localization metrics for different DDOP variants and OTFS pulses, respectively. Finally, Section VII provides numerical results, followed by the conclusion in Section VII. \textcolor{black}{A summary of the key mathematical symbols used in the paper is included in Table I.}

\begin{table}[t]
\color{black}
\caption{Summary of Main Mathematical Symbols.}
\begin{center}
\begin{tabular}{|l|l|}
\hline
\textbf{Symbol}&\textbf{Description}     \\ \hline
\multicolumn{2}{|c|}{\textbf{\textit{ODDM and DDOP Parameters}}}    \\ \hline
\( \mathrm{T}_{\textrm{tot}} = NT \)    & Allocated time domain (TD) resource\\ \hline
\( \mathrm{W}_{\textrm{tot}} = \frac{M}{T} \)   & Allocated frequency domain (FD) resource\\ \hline
$N$    & Number of symbols in the Doppler domain/ \\ 
~   & ~~Number of fine-frequency (Doppler) subcarriers \\\hline
$M$    & Number of symbols in the delay domain/ \\
~   & ~~Number of ODDM symbols \\ \hline
 $\mathcal{F}\triangleq\frac{1}{\mathrm{T}_{\textrm{tot}}}$   & Fine-frequency (Doppler) resolution/ \\  
$~~~=\frac{1}{NT}$   & ~~Subcarrier spacing \\\hline
 $\mathcal{T}\triangleq\frac{1}{\mathrm{W}_{\textrm{tot}}}$    & Fine-time (delay) resolution/ \\
$~~~=\frac{T}{M}$  & ~~Inter symbol duration \\\hline
$x(t)$ & Base-band ODDM signal      \\\hline
$x_m(t)$    & ODDM symbol  \\\hline
$u(t)$ & DD plane orthogonal pulse (DDOP)     \\ \hline
$a(t)$ & Sub-pulse/prototype of the DDOP      \\\hline
$\beta$ & Roll-off factor of the sub-pulse    \\\hline
$T_{g}$ & Duration of the pulse $g(t)$, where $g(t)\in\{a(t),u(t)\}$      \\\hline
$B_{g}$ & Bandwidth of the pulse $g(t)$, where $g(t)\in\{a(t),u(t)\}$      \\ \hline
$D$ & Extension parameter for CP/CS in the general DDOP     \\ \hline
\multicolumn{2}{|c|}{\textbf{\textit{TF Localization Metrics}}}    \\ \hline
$\Delta A$ & TF area (TFA)/Dispersion product \\\hline
$\Delta T$ & Time dispersion/Pulse's effective duration \\ \hline
$\Delta F$  & Frequency dispersion/Pulse's effective bandwidth \\  \hline
$\kappa$ & Direction parameter \\  \hline
\end{tabular}\label{tab1}
\end{center}\vspace{-3mm}
\color{black}
\end{table}

\section{Preliminaries}\label{Sec:Prem}



In this section, we first introduce the ODDM modulation and the DDOP. Thereafter, we discuss the metrics used to quantify the TF localization characteristics of a pulse.
Finally, we present the pulses of two other well-established modulation schemes, which will be used as the benchmark in Section \ref{Sec:Dis} to understand the TF localization characteristics of the DDOP.

\subsection{ODDM Modulation and DDOP}

\color{black}
Let us consider the transmission of $MN$ information-bearing symbols within the allocated TD resource \( \mathrm{T}_{\textrm{tot}} = NT \) and FD resource \( \mathrm{W}_{\textrm{tot}} = \frac{M}{T} \).
In ODDM, first, the information-bearing symbols are mapped into the DD domain to obtain  the ODDM frame $X_{\textrm{DD}}[m,n]$,
where $m\in\{0,1, \cdots,M-1\}$ and $n\in\{0,1, \cdots,N-1\}$  denote the fine time (delay) and fine frequency (Doppler) indices, respectively, and $M$ and $N$ denote the total number of symbols in delay and Doppler domains, respectively. 
Then, all the Doppler domain symbols in a given delay bin, i.e.,  $X_{\textrm{DD}}[m,n]$, $\forall n\in\{0,1,\cdots,N-1\}$ for a given $m$, are modulated onto carriers with separation $\mathcal{F}$ and then aggregated to obtain the $m$th ODDM symbol as~\cite{2022_TWC_JH_ODDM}\footnote{\textcolor{black}{We clarify that utilizing the connection between multicarrier modulation and the inverse discrete Fourier transform (IDFT), discovered by Weinstein \textit{et al.}~\cite{1971_Weinstein_OFDM_usingDFT}, multicarrier-based ODDM can be indirectly (digitally) implemented using several digital transformations and a digital-to-analog converter (DAC)/pulse shaper. 
The details of this digital implementation of ODDM can be found in \cite[Section VII-C]{2023_ODDM_TCOM} and \cite[Section II-A]{2024_TCOM_JunTong_ODDMoverPhysicalChannels}.}}
\begin{align} \label{Equ:ODDMsignal2}
x_m(t)&=u(t)\sum_{n=-\frac{N}{2}}^{\frac{N}{2}-1} X_{\mathrm{DD}}[m,[n]_N]e^{j2\pi n \mathcal{F}t}.  
\end{align}
Here, $[.]_N$ denotes the mod $N$ operation and $\mathcal{F}$ denotes the Doppler resolution of ODDM modulation, and it adheres to the relationship $\mathcal{F}\triangleq\frac{1}{\mathrm{T}_{\textrm{tot}}}=\frac{1}{NT}$. 
Next $x_m(t)$, $\forall m\in\{0,1,\cdots,M-1\}$, are staged 
and aggregated to obtain the base-band ODDM signal $x(t)$  as
\begin{align} \label{Equ:ODDMsignal3}
x(t)
&=\sum_{m=0}^{M-1}\sum_{n=-\frac{N}{2}}^{\frac{N}{2}-1} X_{\mathrm{DD}}[m,[n]_N]u(t-m\mathcal{T})e^{j2\pi n \mathcal{F}(t-m\mathcal{T})},
\end{align}
where $\mathcal{T}$ denotes the delay resolution of ODDM modulation, and it adheres to the relationship $\mathcal{T}\triangleq\frac{1}{\mathrm{W}_{\textrm{tot}}}=\frac{T}{M}$.
From \eqref{Equ:ODDMsignal2} and \eqref{Equ:ODDMsignal3}, it is interesting to observe that $N$ and $M$ can also be regarded as the number of subcarriers and the number of multi-carrier symbols in ODDM modulation, respectively. 
\color{black}

\begin{figure}[t]
\centering
\includegraphics[width=1\columnwidth]{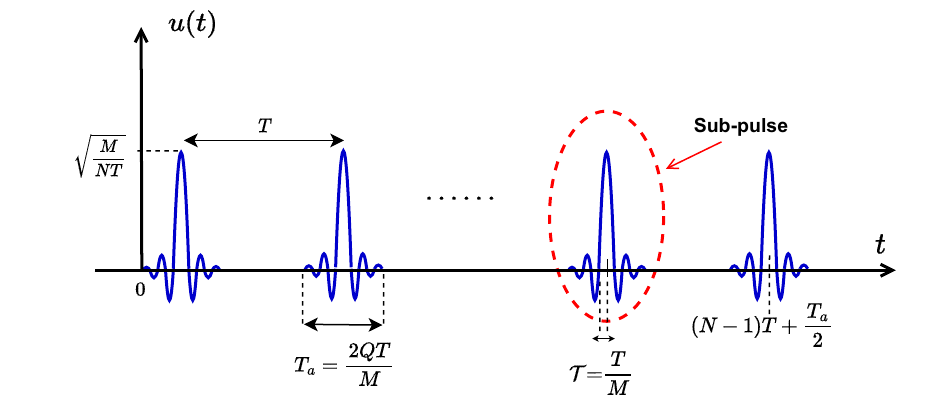}
\caption{Illustration of the DDOP.}\label{Fig:ut}
\end{figure}

At the heart of ODDM modulation is the recently discovered prototype pulse $u(t)$, referred to as the DDOP~\cite{2022_TWC_JH_ODDM}. The DDOP is obtained by concatenating $N$ sub-pulses, $a(t)$, that are spaced apart by the duration $T$, as shown in Fig. \ref{Fig:ut}. 
Mathematically, the DDOP can be expressed as~\cite{2022_TWC_JH_ODDM}
\begin{align}\label{ut}
u(t)=\sum_{n=0}^{N-1}a\left(t-nT-\frac{T_a}{2}\right),
\end{align}
where $a(t)$ denotes the sub-pulse/prototype of the DDOP. The sub-pulse $a(t)$ in \eqref{ut} can be the truncated version of any SRN pulse that is parameterized by (i) the zero ISI interval which is equal to $\mathcal{T}=\frac{T}{M}$, and (ii) the duration constraint $T_{a}=2Q\frac{T}{M}\ll T$ where $Q$ is a positive integer~\cite{2022_TWC_JH_ODDM}.\footnote{\textcolor{black}{
The choice of \(T_{a}\) for DDOP needs to satisfy
$\int_{-T_a/2}^{T_a/2} e^{j2\pi \frac{\tilde{n}}{NT}\bar{t}} a(\bar{t})a^*(\bar{t}-t)\mathrm{d}\bar{t}=\delta(t)$, $\forall|\tilde{n}|\in\{1,\cdots, N-1\},|\tilde{m}|\in\{M-2Q,M-2Q+1,\cdots, M-1\}$,
when cyclic prefix (CP) and cyclic suffix (CS) are not added to the ODDM~[2]. 
We clarify that the simplified criterion \(T_{a} \ll T\) ensures that this satisfaction is achieved.
On another note, recently, a generalized DDOP design was introduced in which CP and CS are added to the DDOP~[7, 8]. For this generalized DDOP, neither this criterion nor the \(T_a \ll T\) criterion is required. The  generalized DDOP design with CP and CS will be discussed in Section V-B of the revised manuscript.}} 

There are several candidates for the SRN pulse in the literature, including the well-known RRC pulse~\cite{Book_GaborAnalysis} and the BTRRC pulse~\cite{2001_ComLet_Normal_BTNP,2004_ComLet_BTRC}. 
Without loss of generality, when the RRC pulse is considered as the SRN pulse. 
the sub-pulse $a(t)$ in \eqref{ut} becomes
\begin{align}\label{at_RRC}
a(t)=
\begin{cases}
\sqrt{\frac{M}{T}\mathcal{E}_{a}}\frac{\sin \left(\pi \frac{Mt}{T}(1-\beta)\right)+\frac{4\beta Mt}{T}~\cos \left( \frac{\pi M t}{T}(1+\beta)\right)}{\frac{\pi Mt}{T}\left(1-\left( \frac{4 \beta Mt}{T}\right)^2\right)},
&\hspace{-3mm}|t|\leq\frac{T_a}{2}, \\
0,&\hspace{-3mm}|t|>\frac{T_a}{2},
\end{cases}
\end{align}
where $\beta$ denotes the roll-off factor which can take values ranging from $0$ to $1$ and $\mathcal{E}_{a}=\frac{1}{N}$ is the energy of the sub-pulse $a(t)$.

\begin{figure}[t]
\centering
\includegraphics[width=1\columnwidth]{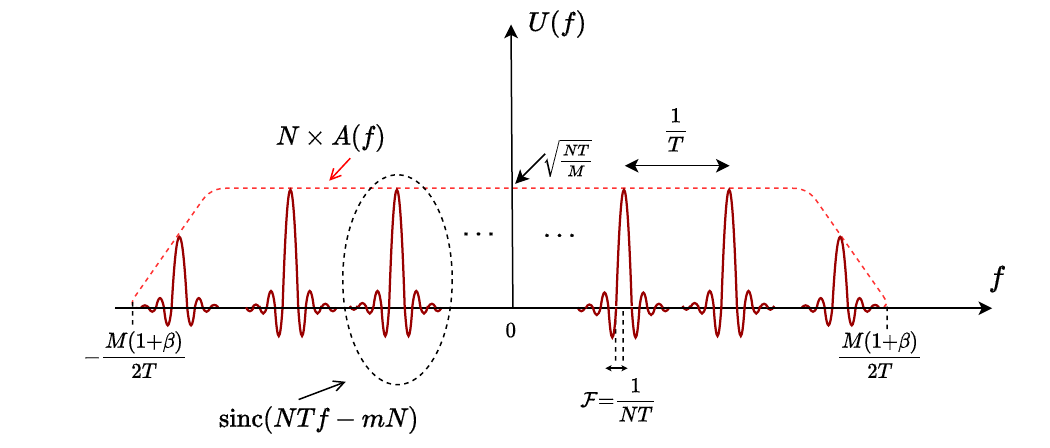}
\caption{The frequency response of the DDOP, where phase terms are ignored for the sake of display.}\label{Fig:Uf}
\end{figure}

\textcolor{black}{As for the frequency response of the DDOP (see Fig. \ref{Fig:Uf}), 
it can be derived as~\cite{2022_GC_JH_ODDMPulse,2023_ODDM_TCOM}}\footnote{\color{black}
As will be discussed later in \eqref{ut_new} in Section V-A, the DDOP $u(t)$ can be expressed in terms of an impulse train, a rectangular TD window, and the sub-pulse $a(t)$~\cite{2022_GC_JH_ODDMPulse,2023_ODDM_TCOM}. 
By taking the Fourier transform on \eqref{ut_new} while applying the time-shifting, convolution, and multiplication properties of the Fourier transform, the $U(f)$ can be derived as in \eqref{Uf}.
More details of this derivation can be found in~\cite[Section VII-A]{2023_ODDM_TCOM}. \color{black}}
\begin{align}\label{Uf}
U(f)=&N e^{-j \pi \left((N-1)T+T_a\right)f} A(f)\notag\\
&\times\sum_{m=-\infty}^{\infty}e^{j \pi (N-1)m}\sinc(NTf-mN),
\end{align}
where $A(f)$ is the frequency response of the chosen $a(t)$. 
\textcolor{black}{For $a(t)$ given in \eqref{at_RRC}, its corresponding $A(f)$ becomes
\begin{align} \label{Af_RRC}
A(f) = \tilde{A}(f) \star \sinc(T_a f),
\end{align}
where \(\tilde{A}(f)\) is the frequency response of the untruncated RRC pulse, and is given by\footnote{\color{black}
We note that since \(A(f)\) in \eqref{Af_RRC} is obtained as the convolution of \(\tilde{A}(f)\) and \(\sinc(T_a f)\), \(A(f)\) exhibits side-lobes extending beyond the spectrum of \(\tilde{A}(f)\), ranging from \(-\frac{M(1+\beta)}{2T}\) to \(\frac{M(1+\beta)}{2T}\). However, we found that these side-lobes are negligibly small. Considering this and for the sake of simplicity, the analytical results in Sections III and V are derived while ignoring the side-lobes of \(A(f)\), effectively treating \(A(f)\) as \(\tilde{A}(f)\). 
\color{black}}
\begin{align}\label{Af_RRC2}
\tilde{A}(f) =
\begin{cases}
\sqrt{\frac{T}{M}\mathcal{E}_{a}},&\hspace{-30mm}0 \leq|f| \leq \frac{M(1-\beta)}{2 T}, \\
\sqrt{\frac{T}{2M}\mathcal{E}_{a}\left(1+\cos \left(\frac{\pi T}{\beta M}\left(|f|-\frac{M(1-\beta)}{2 T}\right)\right)\right)},\\
&\hspace{-30mm}\frac{M(1-\beta)}{2 T} \leq|f| \leq \frac{M(1+\beta)}{2 T},\\
0,&\hspace{-30mm}\textrm{otherwise}.
\end{cases}
\end{align}}

\textcolor{black}{It is interesting to observe from \eqref{Uf} and Fig. \ref{Fig:Uf} that, similar to how the energy of the DDOP is scattered in the TD across multiple sub-pulses $a(t)$ spaced apart by $T$,  its energy is also scattered in the FD across multiple sub-tones $\mathrm{sinc}(NTf)$ spaced apart by $\frac{1}{T}$.
This scattered energy in the FD causes $u(t)$ to occupy a bandwidth of approximately $\frac{M}{T}$. Due to this, the ODDM signals occupy the overall bandwidth of approximately $\frac{M}{T}$,  even though the subcarrier spacing of ODDM modulation is $\mathcal{F}\triangleq\frac{1}{NT}$ and only $N$ subcarriers are available in ODDM.}

\subsection{TF Localization Metrics of a Pulse}

\begin{figure}[t]
\centering
\includegraphics[width=1\columnwidth]{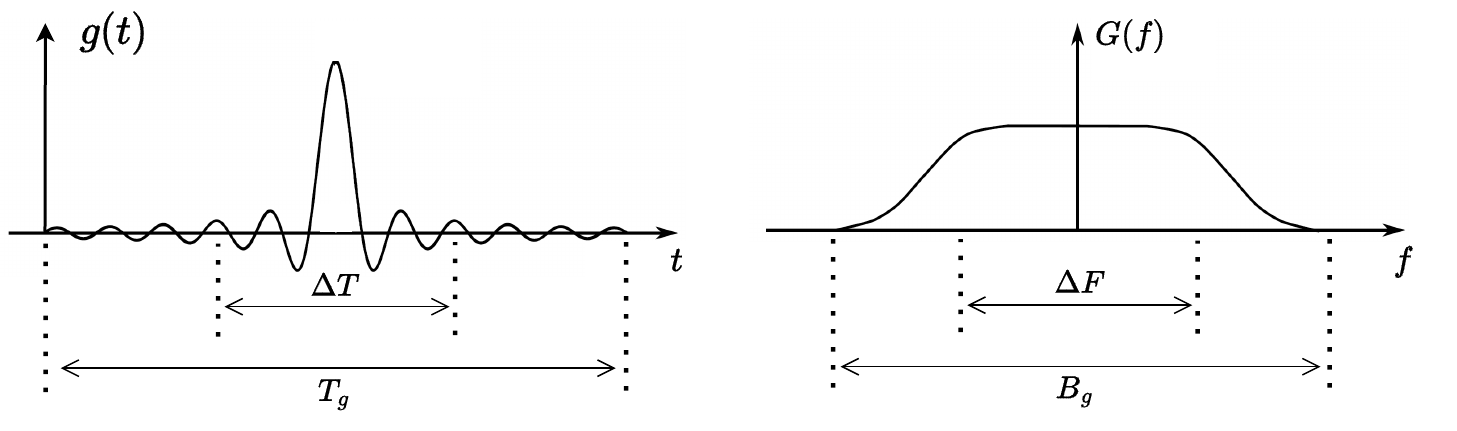}
\caption{Illustration of the metrics $\Delta T$ and $\Delta F$ of a pulse in comparison to its duration $T_g$ and bandwidth $B_g$.}\label{Fig:LocMetrics}
\end{figure}

The TFA, $\Delta A$, time dispersion, $\Delta T$, frequency dispersion, $\Delta F$, and direction parameter, $\kappa$, are classical metrics used in the literature to examine the TF localization characteristics of a pulse~\cite{2023_ODDM_TCOM,2011_SignProcMag_FilterBankMultiCarrier,2017_JSAC_FBMC_forLocalization,2014_ComSurvTut_MCCommunication}. Particularly, they quantify the energy spread of a pulse in the joint TF domain, the TD, the FD, and the TD relative to the FD, respectively. Specifically, $\Delta A$ is defined as the product of $\Delta T$ and $\Delta F$, given by\footnote{To quantify the energy spread of a pulse in the joint TF domain, several studies, e.g.,~\cite{2014_ComSurvTut_MCCommunication,1997__MCSurveyRef85}, opted to utilize the Heisenberg uncertainty parameter, $\epsilon$, instead of the TFA. However, it must be noted that the Heisenberg uncertainly parameter can simply be related to TFA as $\epsilon=\frac{1}{4\pi \Delta A}$.}
\begin{align}\label{Au}
\Delta A = \Delta T\Delta F.
\end{align}
Moreover, $\Delta T$ and $\Delta F$ are defined as the standard deviation of the TD and FD responses of the pulse, respectively, given by~\cite{2014_ComSurvTut_MCCommunication}
\begin{align}\label{DeltaT}
\Delta T =\sqrt{\frac{1}{\mathcal{E}_g}\int_{-\infty}^{\infty}(t-\bar{t}\!~)^2|g(t)|^2 \mathrm{d}t}
\end{align}
and
\begin{align}\label{DeltaF}
\Delta F =\sqrt{\frac{1}{\mathcal{E}_g}\int_{-\infty}^{\infty}(f-\bar{f}\!~)^2|G(f)|^2 \mathrm{d}f},
\end{align}
respectively. In \eqref{DeltaT} and \eqref{DeltaF}, $\mathcal{E}_{g}$ is the energy of the pulse $g(t)$, $G(f)$ is the frequency response of $g(t)$, and
\begin{align}\label{tbatorg}
\bar{t} =\frac{1}{\mathcal{E}_g}\int_{-\infty}^{\infty} t|g(t)|^2 \mathrm{d}t
\end{align}
and
\begin{align}\label{Equ:fbarorg}
\bar{f} =\frac{1}{\mathcal{E}_g}\int_{-\infty}^{\infty} f|G(f)|^2 \mathrm{d}f
\end{align}
represent the mean values of the supports of the pulse in the TD and FD, respectively\footnote{We note that according to the Heisenberg uncertainty principle, time-limited pulses cannot be \textit{strictly} frequency-limited, and frequency-limited pulses cannot be \textit{strictly} time-limited~\cite{Book_TheoryofCommunications_Gabor1946,2023_ODDM_TCOM}. Due to this,
as stated in the Balian-Low Theorem~\cite{Book_TheoryofCommunications_Gabor1946,Book_GaborAnalysis}, either $\Delta T$ or $\Delta F$ of any pulse will take an infinite value. Considering this and the fact that practical pulses are inherently time-limited, in this work, we determine $\Delta F$ of pulses in an essential sense, i.e., \textcolor{black}{$\Delta F$ is determined by ignoring the very small frequency tails that are relatively negligible} ~\cite{2023_ODDM_TCOM}.}. An illustration of $\Delta T$ and $\Delta F$ of a pulse, in comparison with its duration $T_g$ and bandwidth $B_g$, is shown in Fig. \ref{Fig:LocMetrics}. Finally, $\kappa$ is defined as the ratio of $\Delta T$ and $\Delta F$, given by~\cite{2014_ComSurvTut_MCCommunication,1997__MCSurveyRef85}
\begin{align}\label{Equ:kappa}
\kappa=\frac{\Delta T}{\Delta F}.
\end{align}

As a consequence of the Heisenberg uncertainty principle, $\Delta A$ obeys a lower bound known as the \textit{Gabor limit}, given by $\Delta A\geq\frac{1}{4\pi}$. The Gabor limit is attained by the Gaussian pulse~\cite{Book_TheoryofCommunications_Gabor1946,Book_GaborAnalysis}. Typically, \textcolor{black}{a pulse is considered to have minimal energy spread in the joint TF domain (or to be well-localized in the joint TF domain)} if its $\Delta A$ approaches the Gabor limit~\cite{2017_JSAC_FBMC_forLocalization,2014_ComSurvTut_MCCommunication}.

Different from the evaluation of the energy spread of a pulse in the joint TF domain using $\Delta A$, the evaluation of whether a pulse has a minimal energy spread
in the TD, the FD, or/and the TD relative to the FD, using $\Delta T$, $\Delta F$, and $\kappa$, respectively, requires a comparative approach involving another pulse, i.e., a benchmark pulse~\cite{2017_JSAC_FBMC_forLocalization,2015_TFLocalizationforBOTDR,2014_ComSurvTut_MCCommunication}.
Particularly, a pulse is considered to have a lower energy spread in the TD compared to a benchmark pulse when its $\Delta T$ is smaller compared to that of the benchmark pulse. Moreover, a pulse is considered to have a lower energy spread in the FD compared to a benchmark pulse when its $\Delta F$ is smaller compared to that of the benchmark pulse. \textcolor{black}{\textcolor{black}{Furthermore, the amount of stretch of a pulse along the TD relative to the FD is said to be higher compared to a benchmark pulse when its $\kappa$ is larger compared to the benchmark pulse~\cite{2017_JSAC_FBMC_forLocalization,2015_TFLocalizationforBOTDR,2014_ComSurvTut_MCCommunication}.}}

\subsection{TDM and FDM Schemes}

As aforementioned, the evaluation of whether a pulse has a minimal energy spread in the TD, the FD, or/and the TD relative to the FD necessitates a comparative approach involving a benchmark pulse. In this subsection, we consider the pulses
used in two well-established modulation schemes, time division multiplexing (TDM) modulation (a.k.a the single-carrier modulation) and frequency division multiplexing (FDM) modulation. For consistency and fair comparison, we consider that TDM and FDM modulation schemes aim to transmit the same number of symbols using the same TD and FD resources as those of ODDM modulation, i.e., $MN$ symbols using the TD and FD resources corresponding to $\mathrm{T}_{\textrm{tot}}=NT$ and $\mathrm{W}_{\textrm{tot}}=\frac{M}{T}$, respectively.

\subsubsection{TDM Scheme}

In TDM schemes, information is carried by pulses that are narrow in the TD~\cite{2004_FundamentalsofWC_DavidTse}. Thus, its prototype pulse $u_{\mathrm{TDM}}(t)$ is the truncated version of the SRN pulse that is parameterized by the zero ISI interval\footnote{For a fair comparison between DDOP and the pulse used in the TDM scheme, and considering that practical pulses are inherently time-limited, we set the duration of the pulse used in the TDM scheme to be $T_{a}=2Q\frac{T}{M}\ll T$.} which is equal to the time resolution of the TDM scheme  $\mathcal{T}_{\textrm{TDM}}=\frac{\mathrm{T}_{\textrm{tot}}}{MN}=\frac{T}{M}$. In other words, the pulse used in the TDM scheme is simply the sub-pulse used in the DDOP. Similar to the sub-pulse in the DDOP, we consider the RRC pulse as the SRN pulse. Thus, the pulse used in the TDM scheme $u_{\mathrm{TDM}}(t)$ is $u_{\mathrm{TDM}}(t)=a\left(t-\frac{T_a}{2}\right)$ and its frequency response $U_{\mathrm{TDM}}(f)$ is $U_{\mathrm{TDM}}(f)=e^{-j\pi fT_a}A(f)$, where $a(t)$ and $A(f)$ are given by \eqref{at_RRC} and \eqref{Af_RRC}, respectively, with $\mathcal{E}_a=1$. We clarify that the RRC pulses modulated by spreading codes/sequences were adopted in code-division multiple access (CDMA) techniques for the third-generation (3G) communication systems~\cite{200_VTC_CDMARaisedCosine}.

\subsubsection{FDM Scheme}

In FDM schemes, information is carried by pulses that have a narrow frequency response~\cite{2004_FundamentalsofWC_DavidTse}. Thus, its prototype pulse is the rectangular pulse $u_{\mathrm{FDM}}(t)=\frac{1}{\sqrt{NT}}\Pi_{NT}\left(t-\frac{NT}{2}\right)$.
Note that the frequency response of $u_{\mathrm{FDM}}(t)$ is given by $U_{\mathrm{FDM}}(f)=\sqrt{NT} e^{-j\pi fNT} \sinc(NTf)$, and its zero ICI interval is equal to the frequency resolution of the FDM scheme $\mathcal{F}_{\textrm{FDM}}=\frac{\mathrm{W}_{\textrm{tot}}}{MN}=\frac{1}{NT}$. We clarify that with OOBE suppression approaches (e.g., the weighted overlap add (WOLA)), rectangular pulses are adopted in OFDM techniques for the fourth- and fifth-generation (4G/5G) communication systems~\cite{4GwithOFDM,2018_ComStandardMag_OFDMNumerology}.

\section{Analysis of TF Localization Metrics}\label{Sec:DeltaTDerivation}

In this section, we first derive the TF localization metrics of the DDOP. Then, we derive the TF localization metrics of the pulses used in TDM and FDM schemes for the sake of benchmark comparison.

\subsection{TF Localization Metrics of DDOP}\label{Sec:DeltaTFofDDOPDerivation}

We first start the derivation of $\Delta T$ of the DDOP by expanding \eqref{DeltaT} using \eqref{ut} to obtain
\begin{align}\label{DeltaT2}
&\Delta T^2\notag\\
&=\int_{-\infty}^{\infty}t^2|u(t)|^2 \mathrm{d}t-2\bar{t}\int_{-\infty}^{\infty}t|u(t)|^2 \mathrm{d}t+\bar{t}^{2}\int_{-\infty}^{\infty}|u(t)|^2\mathrm{d}t\notag\\
&=\int_{-\infty}^{\infty}t^2|u(t)|^2\mathrm{d}t-\bar{t}^{2}\notag\\
&=\sum_{n=0}^{N-1}\int_{-\infty}^{\infty} t^2\left|a\left(t\!-\!nT\!-\!\frac{T_a}{2}\right)\right|^2 \mathrm{d}t+\ell_{T}-\bar{t}^{2},
\end{align}
where $\ell_{T}$ is given by
\begin{align}
&\ell_{T}\notag\\
&=\sum_{n=0}^{N-1}\sum_{\substack{\dot{n}=0,\\\dot{n}\neq n}}^{N-1}\int_{-\infty}^{\infty}t^2
\left|a\left(t\!-\!nT\!-\!\frac{T_a}{2}\right)
a\left(t\!-\!\dot{n}T\!-\!\frac{T_a}{2}\right)\right|^2\mathrm{d}t.\notag
\end{align}
We note that $\ell_{T}=0$ because when $n\neq\dot{n}$, $a\left(t-nT-\frac{T_a}{2}\right)$ does not overlap with $a\left(t-\dot{n}T-\frac{T_a}{2}\right)$ since the duration of $a(t)$ obeys the condition $T_a\ll T$. Next, we apply Lemma \ref{Lem:t2shiftAll} given in Appendix \ref{App:Lemmas} into the first term in the last equation in \eqref{DeltaT2}. By doing so, we further simplify $\Delta T^2$ as
\begin{subequations}\label{T11}
\begin{alignat}{2}
&\Delta T^2 \notag\\
&=\sum_{n=0}^{N-1}\left(\int_{-\infty}^{\infty} t^2|a(t)|^2 \mathrm{d}t+ \frac{\left(n T+\frac{T_a}{2}\right)^2}{N} \right)-\bar{t}^{2}\notag\\
&=N\int_{-\infty}^{\infty} t^2|a(t)|^2 \mathrm{d}t+\frac{T^2}{N} \sum_{n=0}^{N-1} n^2+\frac{TT_a}{N} \sum_{n=0}^{N-1} n+\frac{T^2_a}{4}-\bar{t}^{2}\label{T11_a}\\
&=N\int_{-\infty}^{\infty} t^2|a(t)|^2 \mathrm{d}t+T(N-1)\left(\frac{T(2N-1)}{6}+\frac{T_a}{2}\right)\notag\\
&~~~+\frac{T^2_a}{4}-\bar{t}^{2},\label{T11_b}
\end{alignat}
\end{subequations}
where \eqref{T11_b} is obtained by using the geometric progression formula~\cite[Eq (0.112)]{IntegralBook} in \eqref{T11_a}. Next, to determine $\bar{t}$ in \eqref{T11_b}, we expand \eqref{tbatorg} using \eqref{ut} while following the steps similar to those used to obtain \eqref{T11_b} from \eqref{DeltaT2}. By doing so, we obtain $\bar{t}$ in \eqref{T11_b} as
\begin{align}\label{Equ:tbar}
\bar{t}=\frac{T(N-1)+T_a}{2}.
\end{align}
Thereafter, substituting \eqref{Equ:tbar} into \eqref{T11_b} and further simplifying it while considering $N$ is sufficiently large, we obtain
\begin{align}\label{T11Extra}
\Delta T^2 \approx\underbrace{N\int_{-\infty}^{\infty} t^2|a(t)|^2 \mathrm{d}t}_{\Delta T^2_{1}}+\underbrace{\frac{N^2T^2}{12}}_{\Delta T^2_{2}}.
\end{align}

Due to the complex nature of the expressions for $a(t)$ in \eqref{at_RRC}, it is extremely difficult, if not impossible, to compute the integral in $\Delta T^2_{1}$ in \eqref{T11Extra}. However, we note that, on the one hand, $\Delta T_{1}$ in \eqref{T11Extra} represents $\Delta T$ of the sub-pulse $a(t)$; on the other hand, $\Delta T$ of any pulse is upper bounded by its duration since $\Delta T$ is simply the standard deviation of the TD representation of the pulse~\cite{2014_ComSurvTut_MCCommunication,1997_Hass_TFLocalization}. Considering these, we
obtain
\begin{align}\label{T111_A}
\Delta T_1 \leq T_a.
\end{align}
Next, we substitute \eqref{T111_A} into \eqref{T11Extra} and further simplify it while considering $T_a\ll T$ and $N$ is sufficiently large. By doing so, we finally obtain $\Delta T$ of the DDOP as\footnote{\textcolor{black}{We note that although $\Delta T$ in \eqref{DeltaT4} is derived for \(T_a \ll T\), it can also be applied for any choice of \(T_a\), including \(T_a < T\), \(T_a = T\), or \(T_a > T\). This possibility arises because, even when \(T_a\) does not strictly satisfy \(T_a\ll T\) to allow the direct simplification from \eqref{DeltaT2} to (\ref{T11}a), the energy of sub-pulses in DDOP is primarily concentrated within the duration of their initial zero crossings. This allows for the simplification from \eqref{DeltaT2} to (\ref{T11}a)  for all $T_a\lesseqgtr T$ with minimal approximation errors.}}
\begin{align}\label{DeltaT4}
\Delta T \approx \frac{NT}{\sqrt{12}}.
\end{align}

We next derive $\Delta F$ of the DDOP. We start by expanding \eqref{DeltaF} using \eqref{Uf} to obtain
\begin{align}\label{DeltaF2}
\Delta F^2=&\int_{-\infty}^{\infty}f^2|U(f)|^2 \mathrm{d}f-2\bar{f}\int_{-\infty}^{\infty}f|U(f)|^2 \mathrm{d}f\notag\\
&+\bar{f}^{2}\int_{-\infty}^{\infty}|U(f)|^2 \mathrm{d}f\notag\\
=&\int_{-\infty}^{\infty}f^2|U(f)|^2 \mathrm{d}f-\bar{f}^{2}\notag\\
=&N^2\sum_{m=-\infty}^{\infty}\int_{-\infty}^{\infty} f^2|A(f)\sinc(NTf-mN)|^2 \mathrm{d}f\notag\\
&+\ell_{F}-\bar{f}^{2},
\end{align}
where $\ell_{F}$ is given by
\begin{align}
\ell_{F}=&N^2\sum_{m=-\infty}^{\infty}\sum_{\substack{\dot{m}=-\infty,\\\dot{m}\neq m}}^{\infty}\int_{-\infty}^{\infty} f^2|A(f)|^2\Big|\sinc(NTf-mN)\notag\\
&\times\sinc(NTf-\dot{m}N)\Big|\mathrm{d}f.\notag
\end{align}
For sufficiently large $N$, we find that $\ell_{F}\approx0$, since the energy of $\sinc(NTf)$ concentrated beyond its $\frac{N}{2}$th zero-crossing point is negligible. Also, $\bar{f}$ in \eqref{DeltaF2} can be derived as zero based on the even symmetry of $|U(f)|$.

Next, we note that it is extremely difficult to analytically compute the integral in the first term in the last equation in \eqref{DeltaF2} for $A(f)$ given in \eqref{Af_RRC}. Considering this, we first approximate \eqref{DeltaF2} as
\begin{align}\label{DeltaF3}
\Delta F^2 \approx N^2\!\!\sum_{m=-\infty}^{\infty}\left|A\left(\frac{m}{T}\right)\right|^2 \int_{-\infty}^{\infty} f^2|\sinc(NTf-mN)|^2 \mathrm{d}f.
\end{align}
We clarify that the impact of this approximation is marginal due to facts that (i) the frequency range where a considerable amount of energy of $\sinc(NTf)$ is concentrated is relatively small as compared to the span of $A(f)$ and (ii) $A(f)$ varies only within $\frac{M(1-\beta)}{2 T} \leq|f| \leq \frac{M(1+\beta)}{2 T}$.
Next, applying Lemma \ref{Lem:t2shiftAll} given in Appendix \ref{App:Lemmas} into \eqref{DeltaF3} and then simplifying it, we obtain
\begin{subequations}\label{DeltaF4}
\begin{align}
\Delta F^2 \approx & N^2\sum_{m=-\infty}^{\infty}\left|A\left(\frac{m}{T}\right)\right|^2 \int_{-\infty}^{\infty} f^2| \sinc(NTf)|^2 \mathrm{d}t\notag\\
&+N^2\sum_{m=-\infty}^{\infty}\left|A\left(\frac{m}{T}\right)\right|^2 \left(\frac{mN}{NT}\right)^2 \frac{1}{NT}\label{DeltaF4_a}\\
\approx&N\sum_{m=-\infty}^{\infty}\left(\frac{m}{T}\right)^2
\left|A\left(\frac{m}{T}\right)\right|^2\frac{1}{T}\notag\\
&+\frac{\mathcal{K}}{\pi^2 T^2N}\sum_{m=-\infty}^{\infty}\left|A\left(\frac{m}{T}\right)\right|^2\frac{1}{T}.\label{DeltaF4_b}
\end{align}
\end{subequations}
We note that \eqref{DeltaF4_b} is obtained by simplifying the first term in \eqref{DeltaF4_a} while considering that the energy concentration of $\sinc(NTf)$ beyond its $\mathcal{K}$th zero-crossing point is negligible. We then use the mathematical identity $\lim_{\Delta \rho\rightarrow 0} \sum_{i\in \mathbb{Z}}\mathcal{X}(i\Delta \rho)\Delta \rho=\int_{-\infty}^{\infty}\mathcal{X}(\rho)d\rho$ into \eqref{DeltaF4_b}, which allows us to approximate \eqref{DeltaF4_b} as
\begin{align}\label{DeltaF45}
\Delta F^2 \approx \underbrace{N\int_{-\infty}^{\infty}f^2\left|A(f)\right|^2 \mathrm{d}f}_{\Delta F^2_{1}}+\underbrace{\frac{\mathcal{K}}{\pi^2 T^2N}\int_{-\infty}^{\infty}\left|A(f)\right|^2 \mathrm{d}f}_{\Delta F^2_{2}}.
\end{align}
As for $\Delta F_{1}^2$ in \eqref{DeltaF45}, it can be derived using $A(f)$ in \eqref{Af_RRC} as \eqref{DeltaFXX}, which is given at the end of next page.
\begin{figure*}[!b]
\hrulefill
\normalsize 
\begin{alignat}{2}\label{DeltaFXX}
\Delta F_{1}^2&=\frac{2T}{M}\int_{0}^{\frac{M(1-\beta)}{2 T}} f^2\textrm{d}f+\frac{T}{M}\int_{\frac{M(1-\beta)}{2 T}}^{\frac{M(1+\beta)}{2 T}} f^2\left(1+\cos \left(\frac{\pi T}{\beta M}\left(f-\frac{M(1-\beta)}{2 T}\right)\right)\right)\textrm{d}f\notag\\
&=\frac{2T}{3M}\left(\frac{M(1-\beta)}{2T}\right)^3
+\frac{M^2\beta(\pi^2\beta^2-24\beta+3\pi^2)}{12\pi^2T^2}\notag\\
&=\frac{M^2}{12T^2}+\frac{(\pi^2-8)M^2\beta^2}{4\pi^2T^2}.
\end{alignat}
\end{figure*}
Then, we find that $\Delta F^2_{2}$ in \eqref{DeltaF45} is $\Delta F^2_{2}=\frac{\mathcal{K}}{\pi^2 T^2N^2}\ll \Delta F^2_{1}$. Considering this, we obtain $\Delta F$ of the DDOP as
\begin{align}\label{DeltaF5}
\Delta F\approx\frac{M}{T}\sqrt{\frac{1}{12}+\frac{(\pi^2-8)\beta^2}{4\pi^2}}.
\end{align}

\noindent
We note that the derivations of $\Delta F$ in \eqref{DeltaF5} and $\Delta T$ in \eqref{DeltaT4} involve several approximations and considerations, \textcolor{black}{including the negligible energy of $\sinc(NTf)$ beyond its $\frac{N}{2}$th zero-crossing point, and sufficiently large $M$ and $N$.} 
 Despite so, in Section \ref{Sec:Num} we will show the accuracy of our analysis through numerical results for a wide range of system parameters, thereby validating the accuracy of our derivations.

Finally, based on the results derived in \eqref{DeltaT4} and \eqref{DeltaF5}, we obtain $\Delta A$ and $\kappa$ of the DDOP. These results, along with those in \eqref{DeltaT4} and \eqref{DeltaF5}, are presented in the following theorem.
\begin{theorem}\label{Thr:IORCPOTFS-VCP-ECU}
The TF localization metrics of the DDOP are given by
\begin{subequations}
\begin{align}
\Delta A_{\textrm{DDOP}}&\approx\frac{MN}{12}\sqrt{1+\frac{3(\pi^2-8)\beta^2}{\pi^2}},\label{DeltaA_DDOP}\\
\Delta T_{\textrm{DDOP}} &\approx\frac{NT}{\sqrt{12}}, \label{DeltaT_DDOP}\\
\Delta F_{\textrm{DDOP}} &\approx\frac{M}{T}\sqrt{\frac{1}{12}+\frac{(\pi^2-8)\beta^2}{4\pi^2}},\label{DeltaF_DDOP}\\
\kappa_{\textrm{DDOP}}&\approx\frac{NT^2}{M}\sqrt{\frac{\pi^2}{\pi^2+3(\pi^2-8)\beta^2}}\label{k_DDOP}.
\end{align}
\end{subequations}
\end{theorem}

Prior to interpreting the results in \emph{Theorem \ref{Thr:IORCPOTFS-VCP-ECU}}, in the next subsection, we present the TF localization metrics of the pulses used in TDM and FDM schemes. These metrics will be used as the benchmark in Section \ref{Sec:Dis} to gain a deep understanding of the results in \emph{Theorem \ref{Thr:IORCPOTFS-VCP-ECU}}.

\subsection{TF Localization Metrics of Pulses used in TDM and FDM Schemes}

\subsubsection{TDM Scheme}

By directly calculating the integrals in \eqref{DeltaT}--\eqref{Equ:fbarorg} while using \eqref{at_RRC} and \eqref{Af_RRC} in this calculation, the TF localization metrics of the pulse used in the TDM scheme can be obtained for sufficiently large $M$ as\footnote{When determining $\Delta T$ for the pulse used in the TDM scheme, the challenge is similar to that faced when computing $\Delta T_{1}$ in \eqref{T11Extra}. Thus, only an upper bound is provided for $\Delta T$ of the pulse used in the TDM scheme.}
\begin{subequations}\label{Deltas_TDM}
\begin{align}
\Delta A_{\textrm{TDM}}&\approx\frac{1}{\pi}\sqrt{\frac{Q}{12}+\frac{Q(\pi^2-8)\beta^2}{4\pi^2}},\label{DeltaA_TDM}\\
\Delta T_{\textrm{TDM}} &\approx\frac{T\sqrt{Q}}{M\pi}, \label{DeltaT_TDM}\\
\Delta F_{\textrm{TDM}} &\approx\frac{M}{T}\sqrt{\frac{1}{12}+\frac{(\pi^2-8)\beta^2}{4\pi^2}},\label{DeltaF_TDM}\\
\kappa_{\textrm{TDM}}&\approx\frac{T^2}{M^2}\sqrt{\frac{12Q}{\pi^2+3(\pi^2-8)\beta^2}}\label{k_TDM}.
\end{align}
\end{subequations}

\subsubsection{FDM Scheme}

By directly calculating the integrals in \eqref{DeltaT}--\eqref{Equ:fbarorg} while using $u_{\mathrm{FDM}}(t)$ and $U_{\mathrm{FDM}}(f)$ in this calculation, the TF localization metrics of the pulse used in the FDM scheme can be obtained for sufficiently large $N$ as
\begin{subequations}\label{Deltas_FDM}
\begin{align}
\Delta A_{\textrm{FDM}}&\approx\frac{1}{\pi}\sqrt{\frac{\mathcal{K}}{12}},\label{DeltaA_FDM}\\
\Delta T_{\textrm{FDM}} &\approx\frac{NT}{\sqrt{12}},\label{DeltaT_FDM}\\
\Delta F_{\textrm{FDM}} &\approx\frac{\sqrt{\mathcal{K}}}{NT\pi},\label{DeltaF_FDM}\\
\kappa_{\textrm{FDM}}&\approx\frac{N^2T^2\pi}{\sqrt{12\mathcal{K}}}.\label{k_FDM}
\end{align}
\end{subequations}

\section{Discussion and Remarks}\label{Sec:DiscandRemarks}

In this section, we first discuss the TF localization characteristics of the DDOP based on the results derived in \emph{Theorem \ref{Thr:IORCPOTFS-VCP-ECU}}.  Thereafter, we present several remarks related the characteristics of the DDOP. Finally, we discuss the potential advantages brought by the energy spread of the DDOP.

\subsection{Discussion}\label{Sec:Dis}

By observing \eqref{DeltaA_DDOP} in \emph{Theorem \ref{Thr:IORCPOTFS-VCP-ECU}}, we find that $\Delta A$ of the DDOP does not reach the Gabor limit of $\frac{1}{4\pi}$. This indicates that \textcolor{black}{the energy spread of DDOP in the joint TF domain is high}
Moreover, it is noteworthy to compare \eqref{DeltaA_DDOP} in \emph{Theorem \ref{Thr:IORCPOTFS-VCP-ECU}} with \eqref{DeltaA_TDM} and \eqref{DeltaA_FDM}. This comparison reveals that the energy of the DDOP in the joint TF domain is more widely spread compared to those of the pulses used in TDM and FDM schemes.

By comparing \eqref{DeltaT_DDOP} in \emph{Theorem \ref{Thr:IORCPOTFS-VCP-ECU}} with \eqref{DeltaT_TDM}, we observe that $\Delta T$ of the DDOP is larger compared to that of the pulse used in the TDM scheme. This indicates that the energy of the DDOP in the TD is more widely spread than that of the pulse used in the TDM scheme. By comparing \eqref{DeltaT_DDOP} in \emph{Theorem \ref{Thr:IORCPOTFS-VCP-ECU}} with \eqref{DeltaT_FDM}, we observe that $\Delta T$ of the DDOP is comparable to that of the pulse used in the FDM scheme.

Likewise, by comparing \eqref{DeltaF_DDOP} in \emph{Theorem \ref{Thr:IORCPOTFS-VCP-ECU}} with \eqref{DeltaF_FDM}, we observe that
$\Delta F$ of the DDOP is larger compared to that of the pulse used in the FDM scheme. This indicates that the energy of the DDOP in the FD is more widely spread than that of the pulse used in the FDM scheme. By comparing \eqref{DeltaF_DDOP} in \emph{Theorem \ref{Thr:IORCPOTFS-VCP-ECU}} with \eqref{DeltaF_TDM}, we observe that
$\Delta F$ of the DDOP is comparable to that of the pulse used in the TDM scheme.

Lastly, by comparing $\kappa$ in \eqref{k_DDOP} in \emph{Theorem \ref{Thr:IORCPOTFS-VCP-ECU}} with those in \eqref{k_TDM} and \eqref{k_FDM}, we observe that the amount of stretch of the DDOP along the TD relative to the FD is higher than that of the pulse used in the TDM scheme, but lower than that of the pulse used in the FDM scheme.

\begin{figure*}[t]
\centering
\includegraphics[width=1.5\columnwidth]{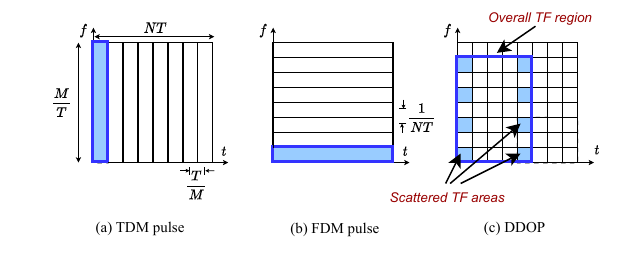}
\caption{Simplified TF occupancy of the DDOP, the pulse used in the TDM scheme, and the pulse used in the FDM scheme.}\label{Fig:Loc_Compare}
\end{figure*}

\subsubsection{Simplified Schematic Illustration}\label{Sec:Schematic1}

We provide a simplified schematic illustration in Fig. \ref{Fig:Loc_Compare} to clearly understand the reasons behind the relatively high $\Delta A$ of the DDOP. This figure demonstrates the specific TF regions where the energy of the DDOP and the pulses used in TDM and FDM schemes are concentrated within a TF region bounded by TD and FD resources of $\mathrm{T}_{\textrm{tot}}=NT$ and $\mathrm{W}_{\textrm{tot}}=\frac{M}{T}$, respectively, with $M=4$ and $N=2$~\cite{2023_ODDM_TCOM}.\footnote{For the sake of simplicity, Fig. \ref{Fig:Loc_Compare} only displays the TF regions where the energy of pulses are predominantly concentrated~\cite{2023_ODDM_TCOM}.}

For the pulses used in TDM and FDM schemes, their energy is concentrated within a single non-scattered region in the TF plane (see Figs. \ref{Fig:Loc_Compare}(a) and \ref{Fig:Loc_Compare} (b)). As for the dimensions of this non-scattered region, one dimension is very low and the other dimension tends to be very large as the two dimensions exhibit an inverse relationship as per the Heisenberg uncertainty principle~\cite{Book_TheoryofCommunications_Gabor1946}. Consequently, in the case of the pulse used in the TDM scheme,  $\Delta T$ is relatively small and $\Delta F$ is relatively high, thereby leading to a relatively small $\Delta A$. In the case of the pulse used in the FDM scheme, $\Delta T$ is relatively high, and $\Delta F$ is relatively small, thereby leading to a relatively small $\Delta A$.

In contrast, as shown in Fig. \ref{Fig:ut}, the energy of the DDOP is scattered in the TD across $N$ sub-pulses $a(t)$. Also, as shown in Fig. \ref{Fig:Uf}, its energy is scattered in the FD across approximately $M$ sub-tones $\sinc(NTf)$.
Due to these, the TF region occupied by the DDOP pulse in the TF plane consists of approximately $MN$ \textit{small scattered TF areas} that are located far apart (see Fig. \ref{Fig:Loc_Compare}(c))~\cite{2023_ODDM_TCOM}. Hence, the energy spread of the DDOP is governed by the dimensions of the \textit{overall TF region} that encompasses all the small scattered TF areas (see the region enclosed by the blue solid line in Fig. \ref{Fig:Loc_Compare}(c)). For this overall TF region, both of its dimensions are relatively large, leading to relatively high $\Delta T$ and $\Delta F$. Consequently, $\Delta A$ of the DDOP is relatively high.

\subsection{Remarks}\label{Sec:Remarks}


\begin{remark}\label{Rem:HUP}
\color{black}
In the realm of DD domain modulation schemes, there is a prevailing notion that pulses well-suited for such schemes may not exist in practice as the delay and Doppler resolutions of such schemes are extremely small,  which result in a joint TF resolution (JTFR) of less than one~\cite{2017_WCNC_OTFS_Haddani,2022_GC_JH_ODDMPulse}. 
However, \eqref{DeltaA_DDOP} in \emph{Theorem \ref{Thr:IORCPOTFS-VCP-ECU}} reveals that the TFA $\Delta A$ of the DDOP exceeds the Gabor limit of $\frac{1}{4\pi}$. This indicates that the DDOP 
does not violate the theoretical limits imposed by the Heisenberg uncertainty principle, thereby confirming the practical existence of pulses well-suited for DD domain modulation schemes~\cite{2003_TFLocalizationMultiCarrier}.
\color{black}

\end{remark}

\begin{remark}\label{Rem:A}
It is important to note that approximately $MN$ small scattered TF areas constitute the DDOP in the joint TF domain. These small scattered TF areas neither overlap nor are located in close proximity to each other. By examining each of these small TF areas separately, we find that their TFA values are on the order of $\frac{1}{4\pi M N}$, which falls bellow the Gabor limit of $\frac{1}{4\pi}$~\cite{Book_TheoryofCommunications_Gabor1946,Book_GaborAnalysis}. This suggests that, when considered in isolation, these small scattered TF areas seem to exhibit extremely low energy spread in the joint TF domain. \textcolor{black}{This implies that the \textit{DDOP behaves locally like a well-localized pulse in the joint TF domain (as well as in the TD and FD), even though it is not globally well-localized in the joint TF domain, the TD, and the FD}.} On another note, it is important to highlight that generating pulses that have their TF occupancy as one of these small scattered TF areas is not feasible, since this attempt would violate the Heisenberg uncertainty principle.
\end{remark}

\begin{figure}[t]
\centering
\includegraphics[width=0.9\columnwidth]{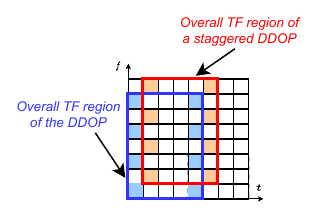}
\caption{Illustration of the overall TF regions of the DDOP and a staggered DDOP. }\label{Fig:DoF} 
\end{figure}

\begin{remark}\label{Rem:B}
As highlighted in~\cite[Section VII-A-2]{2023_ODDM_TCOM}, between the small scattered TF areas that constitute the DDOP in the joint TF domain, the orthogonality w.r.t. the fine time resolution $\mathcal{T}{=}\frac{T}{M}$ and fine frequency resolution $\mathcal{F}{=}\frac{1}{NT}$ is achieved. 
This enables the 
orthogonal staggering of $MN$ DDOPs in the TF plane, while allowing for significant overlap of their overall TF regions. A simplified illustration is given in Fig.~\ref{Fig:DoF}.
This leads to the degree of freedom (DoF) of ODDM modulation to be $MN$ for the TF region bounded by TD and FD resources of $\mathrm{T}_{\textrm{tot}}=NT$ and $\mathrm{W}_{\textrm{tot}}=\frac{M}{T}$, respectively~\cite{2023_ODDM_TCOM}. We clarify that the DoF of a modulation scheme corresponds to its ability to best utilize the available TF resources~\cite{2004_FundamentalsofWC_DavidTse}.
Thus, it is interesting to find that for the TF region bounded by the same TD and FD resources, both TDM and FDM schemes yield the same DoF of $MN$. 
This 
equivalence in DoF is notable,
particularly when considering that $\Delta A$ of the DDOP is relatively high and those of the pulses used in TDM and FDM schemes are relatively low.
\end{remark}

\subsection{Potentials of DDOP}

\subsubsection{Harnessing Channel Diversity with DDOP}

\color{black}
For a modulation scheme to harness the time (Doppler) diversity of a channel, its pulse needs to exhibit a wide energy spread in the TD, and to harness the frequency (delay) diversity, the pulse needs to have a wide energy spread in the FD. Pulses in TDM and FDM schemes display a wide energy spread in either the TD or FD, allowing them to exploit either the time or frequency diversity of the channel, but not both. In contrast, the wide energy spread of the DDOP in both TD and FD enables it to effectively interact with the sparse DD domain representation of doubly selective channels, thereby exploiting both time and frequency diversities of the channel~\cite{2021_WCM_JH_OTFS}.
\color{black}

\subsubsection{Sensing with DDOP}

\textcolor{black}{As highlighted in Remark \ref{Rem:HUP},  even though the DDOP globally has a wide energy spread, it locally behaves like a pulse or a tone with a narrow energy spread in the TD, FD, and joint TF domain. This local narrow energy spread behavior in the TD and FD makes it well-suited for sensing applications that require fine resolutions~\cite{2021_WCM_JH_OTFS,2023_ODDM_TCOM,2020_TWC_Lorenzo_OTFSforJSAC,2022_ISAC_Survey}.}
In particular, the goal of sensing is to accurately estimate \textit{both the range and the velocity of an object}~\cite{2021_WCM_JH_OTFS,2023_ODDM_TCOM}. The range estimation relies on the delay of the backscattered pulse, while the velocity estimation relies on the Doppler shift (or phase variation) of the backscattered pulse. It is important to note that estimating the range of the object with fine time resolution necessitates the backscattered pulse to exhibit a \textcolor{black}{narrow energy spread in the TD}. On the other hand, estimating the velocity of the object with fine frequency resolution necessitates the backscattered pulse to exhibit a \textcolor{black}{narrow energy spread in the FD.}

\textcolor{black}{When using pulses like those in TDM and FDM schemes for sensing, either the range or the velocity of the object can be estimated with fine resolution, but not both. This limitation arises because these pulses exhibit narrow energy spread in only one domain, either the TD or the FD.}  
In contrast to pulses used in TDM and FDM schemes, DDOP enables the \emph{simultaneous estimation} of \emph{both the range and the velocity of an object} with fine time and frequency resolutions, \textcolor{black}{since it behaves locally like a pulse or a tone with narrow energy spread in both TD and FD.} This makes \textcolor{black}{ODDM, and in general all DD domain modulation schemes,} highly suitable for sensing applications that demand fine resolutions~\cite{2024_RISAidedHMC,2020_TWC_Lorenzo_OTFSforJSAC,2022_ISAC_Survey}.

\section{TF localization Metrics of Different Variants of DDOP}\label{Sec:OtherDDOP}

In this section, we first present the relationship between $\Delta T$ and $\Delta F$ of the DDOP and \textcolor{black}{those of the envelope functions of its TD and FD representations.} 
Based on this relationship, we then determine the TF localization metrics of (i) the generalized design of the DDOP and (ii) the DDOP when its sub-pulse is not the RRC pulse.

\subsection{Characterizing $\Delta T$ and $\Delta F$ of the DDOP based on the Envelope Functions of its TD and FD
Representations}\label{Sec:WindowandFilter}

We note that any given DDOP can be obtained by sending an impulse train $\dot{u}(t)=\sum_{n=-\infty}^{\infty}\delta(t-nT)$ through (i) a rectangular TD window $b(t)=\Pi_{NT}\left(t-\frac{(N-1)T}{2}\right)$ and (ii) a filter having the impulse response of the sub-pulse~\cite{2022_GC_JH_ODDMPulse,2023_ODDM_TCOM}. Based on this, the DDOP can be expressed as
\begin{align} \label{ut_new}
u\left(t+\frac{T_a}{2}\right)= \left(\dot{u}(t)\times b(t)\right)\star a(t),
\end{align}
where $\star$ denotes the linear convolution operation. 
\textcolor{black}{It is interesting to observe that because of the possibility to characterize DDOP as in \eqref{ut_new}, on one hand, DDOP's TD representation implicitly has the envelope function $b(t)$. On the other hand, DDOP's FD representation has an envelope function defined by the frequency response of $a(t)$, $A(f)$.}

By analyzing \eqref{T11Extra}, we find that $\Delta T$ of the DDOP can be fully characterized by using $\Delta T$ of the envelope function that characterizes its TD representation. Particularly, \eqref{T11Extra} indicates that $\Delta T_{\mathrm{DDOP}}^2=\Delta T^2_{1}+\Delta T^2_{2}$. Upon close examination, we find that $\Delta T_{1}$ and $\Delta T_{2}$ are same as $\Delta T$ of $a(t)$ and $b(t)$, respectively, i.e., $\Delta T_{1}=\Delta T_{a(t)}$ and $\Delta T_{2}=\Delta T_{b(t)}$.
However, as explained in Section \ref{Sec:DeltaTFofDDOPDerivation}, since $\Delta T_{1}\ll\Delta T_{2}$, we obtain
$\Delta T_{\mathrm{DDOP}}\approx\Delta T_{b(t)}$. 
\textcolor{black}{Given that $b(t)$ is the envelope function of DDOP's TD representation, it can be inferred 
that the \textit{time dispersion of the DDOP} can be determined based on the \textit{time dispersion of the envelope function of DDOP's TD representation}, i.e., 
\begin{align} \label{Equ:DeltaT_relF}
\Delta T_{\mathrm{DDOP}}\approx\Delta T_{\mathrm{DDOP's~TD~envelope~function}}.
\end{align}}


Similarly, \eqref{DeltaF45} indicates that $\Delta F_{\mathrm{DDOP}}^2=\Delta F^2_{1}+\Delta F^2_{2}$, and $\Delta F_{1}$ and $\Delta F_{1}$ are same as $\Delta F$ of $a(t)$ and $b(t)$, respectively, i.e., $\Delta F_{1}=\Delta F_{a(t)}$ and $\Delta F_{2}=\Delta F_{b(t)}$. Since $\Delta F_{2}\ll\Delta F_{1}$, we find that $\Delta F_{\mathrm{DDOP}}\approx\Delta F_{a(t)}$.
\textcolor{black}{Given that $A(f)$ is the envelope function of DDOP's FD representation, it can be inferred 
that the \textit{frequency dispersion of the DDOP} can be characterized based on the \textit{frequency dispersion of the envelope function of DDOP's FD representation} as
\begin{align} \label{Equ:DeltaF_relF}
\Delta F_{\mathrm{DDOP}}\approx\Delta F_{\mathrm{DDOP's~FD~envelope~function}}.
\end{align}}

\color{black}
\begin{remark}\label{Rem:DDOPTDMFDM}
It is interesting to observe that the envelope function of DDOP's TD representation $b(t)$ is same as the pulse used in the FDM scheme $u_{\mathrm{FDM}}(t)$. This observation along with \eqref{Equ:DeltaT_relF} explains why \(\Delta T\) for DDOP in \eqref{DeltaT_DDOP} is similar to \(\Delta T\) of $u_{\mathrm{FDM}}(t)$ in \eqref{DeltaT_FDM}. 
Likewise, the envelope function of DDOP's FD representation is same as the frequency response of the pulse used in the TDM scheme $u_{\mathrm{TDM}}(t)$. 
This observation along with \eqref{Equ:DeltaF_relF} explains why \(\Delta F\) for DDOP in \eqref{DeltaF_DDOP} is similar to \(\Delta F\) of $u_{\mathrm{TDM}}(t)$ in \eqref{DeltaF_TDM}. 
\color{black}
\end{remark}
\color{black}

\subsection{TF Localization Metrics of the Generalized Design of DDOP}\label{Sec:GDDOP}

We note that the DDOP described in \eqref{ut} is formed by concatenating $N$ sub-pulses, with the duration of each sub-pulse constraint by $T_a\ll T$, to ensure the orthogonality conditions for the DDOP are appropriately met. However,~\cite{2022_GC_JH_ODDMPulse,2023_ODDM_TCOM} recently proposed a generalized design of the DDOP in which the duration constraint $T_a\ll T$ of sub-pulses is released, while introducing cyclic prefix (CP) and cyclic suffix (CS) to the pulse. We refer this generalized design of the DDOP as the \emph{general DDOP}, and express it as the concatenation of $N+2D$ sub-pulses spaced apart by $T$ as
\begin{align}\label{ut_general}
\bar{u}(t)=\sum_{n=-D}^{N-1+D}\bar{a}\left(t-(n+D)T-\frac{T_a}{2}\right),
\end{align}
where $D=\lceil\frac{T_a}{T}\rceil$ with $\lceil\cdot\rceil$ denoting the ceil operation. In \eqref{ut_general}, $\bar{a}(t)$ denotes the sub-pulse without the duration constraint $T_a\ll T$; thus, $T_a$ can be $T_a\lesseqgtr T$.

By following the steps similar to those presented in Section \ref{Sec:DeltaTFofDDOPDerivation}, it is indeed possible to determine the TF localization metrics of the general DDOP. However, for simplicity, leveraging the insights obtained in \eqref{Equ:DeltaT_relF} and \eqref{Equ:DeltaF_relF} in Section \ref{Sec:WindowandFilter}, we determine $\Delta T$ and $\Delta F$ of the general DDOP in terms of $\Delta T$ of the envelope function of its TD representation and $\Delta F$ of the envelope function of its FD representation.

For the general DDOP, a rectangular TD window given by $\bar{b}(t)=\Pi_{(N+2D)T}\left(t-\frac{(N+2D-1)T}{2}\right)$ acts as the envelope function of its TD representation, and the frequency response of $\bar{a}(t)$, given by $\bar{A}(f)$,  acts as the envelope function of its FD representation. Also, $\Delta T$  of $\bar{b}(t)$ and $\Delta F$ of $\bar{a}(t)$ can be derived as $\Delta T \approx\frac{(N+2D)T}{\sqrt{12}}$ and $\Delta F \approx\frac{M}{T}\sqrt{\frac{1}{12}+\frac{(\pi^2-8)\beta^2}{4\pi^2}}$, respectively. Based on these results and leveraging the insights obtained from Section \ref{Sec:WindowandFilter}, we deduce $\Delta T$ and $\Delta F$ of the general DDOP as
\begin{align}\label{DeltaT_GDDOP}
\Delta T_{\textrm{G-DDOP}} &\approx\frac{(N+2D)T}{\sqrt{12}}
\end{align}
and
\begin{align}\label{DeltaF_GDDOP}
\Delta F_{\textrm{G-DDOP}}\approx\frac{M}{T}\sqrt{\frac{1}{12}
+\frac{(\pi^2-8)\beta^2}{4\pi^2}},
\end{align}
respectively. Finally, based on \eqref{DeltaT_GDDOP} and \eqref{DeltaF_GDDOP}, we obtain $\Delta A$ and $\kappa$ of the general DDOP as
\begin{align}\label{DeltaA_GDDOP}
\Delta A_{\textrm{G-DDOP}}\approx\frac{M(N+2D)}{12}
\sqrt{1+\frac{3(\pi^2-8)\beta^2}{\pi^2}}
\end{align}
and
\begin{align}\label{k_GDDOP}
\kappa_{\textrm{G-DDOP}}\approx\frac{(N+2D)T^2}{M}
\sqrt{\frac{\pi^2}{\pi^2+3(\pi^2-8)\beta^2}},
\end{align}
respectively.

\textcolor{black}{
We note that, unlike the original DDOP \(u(t)\) in \eqref{ut}, the general DDOP \(\bar{u}(t)\) in \eqref{ut_general} includes \(2D\) additional  sub-pulses. These additional sub-pulses lead to the dependence of the TF localization metrics for the general DDOP in \eqref{DeltaT_GDDOP}-\eqref{k_GDDOP} on \(D\). Since \(D\) is determined by the sub-pulse duration \(T_a\), the general DDOP's TF localization metrics are influenced by \(T_a\), which is different from the original DDOP, whose TF localization metrics are independent of \(T_a\).}

\subsection{TF Localization Metrics of DDOP with BTRRC Sub-pulse}\label{Sec:DDOP_BTRRC}

We recall that the sub-pulse $a(t)$ in the DDOP can be the truncated version of any SRN pulse~\cite{2022_TWC_JH_ODDM}. However, our previous derivations exclusively considered the RRC pulse as the sub-pulse of the DDOP~\cite{Book_GaborAnalysis}.
Another SRN pulse that can be a possible candidate for the sub-pulse is the better-than RRC (BTRRC), also referred to as the Beaulieu pulse~\cite{2001_ComLet_Normal_BTNP,2004_ComLet_BTRC}. We clarify that the BTRRC pulse was introduced in~\cite{2001_ComLet_Normal_BTNP} as an alternative to the RRC pulse, offering a potential advantage in reducing the error probability in the presence of symbol timing errors. The frequency response of the BTRRC pulse is given by
\begin{align}\label{Af_BTRRC}
A(f)=
\begin{cases}
\sqrt{\frac{T}{M}\mathcal{E}_{a}},
&\hspace{-20mm} 0 \leq|f| \leq \frac{M(1{-}\beta)}{2 T}, \\
\sqrt{\frac{T}{M}\mathcal{E}_{a}e^{-\frac{2\log(2)T}{\beta M}\left(|f|-\frac{M}{2T}(1-\beta)\right)}},\\
&\hspace{-20mm}\frac{M(1{-}\beta)}{2 T} \leq|f| \leq \frac{M}{2 T},\\
\sqrt{\frac{T}{M}\mathcal{E}_{a}\left(1-e^{-\frac{2\log(2)T}{\beta M}\left(\frac{M}{2T}(1+\beta)-|f|\right)}\right)},\\
&\hspace{-20mm}\frac{M}{2 T} \leq|f| \leq \frac{M(1{+}\beta)}{2 T},\\
0, &\hspace{-20mm}\textrm{otherwise}.
\end{cases}
\end{align}
Next, leveraging the insights obtained in \eqref{Equ:DeltaT_relF} and \eqref{Equ:DeltaF_relF} in Section \ref{Sec:WindowandFilter}, we determine TF localization metrics of the DDOP when its sub-pulse $a(t)$ is the BTRRC pulse.

\begin{figure*}[!b]
\normalsize\hrulefill
\begin{alignat}{2}\label{DeltaF_BTRRC}
\Delta F^2&=\frac{2T}{M}\int_{0}^{\frac{M(1-\beta)}{2 T}}f^2\textrm{d}f+\frac{2T}{M} \int_{\frac{M(1-\beta)}{2 T}}^{\frac{M}{2 T}}f^2e^{-\frac{2\log(2)T}{\beta M}\left(f-\frac{M}{2T}(1-\beta)\right)}\textrm{d}f+\frac{2T}{M}\int_{\frac{M}{2 T}}^{\frac{M(1+\beta)}{2 T}}f^2\left(1-e^{-\frac{2\log(2)T}{\beta M}\left(\frac{M}{2T}(1+\beta)-f\right)}\right)\textrm{d}f\notag\\
&=\frac{2T}{3M}\left(\frac{M(1-\beta)}{2 T}\right)^3+\frac{M^2{\beta}\left(2\left(\log\left(2\right)-1\right)^2{\beta}^2-2\log\left(2\right)\left(2\log\left(2\right)-1\right){\beta}+\log^2\left(2\right)\right)}{8T^2\log^3\left(2\right)}\notag\\
&~~~-\frac{M^2{\beta}\left(2\left(\log\left(2\right)-1\right)^2{\beta}^2+2\log\left(2\right)\left(2\log\left(2\right)-1\right){\beta}+\log^2\left(2\right)\right)}{8T^2\log^3\left(2\right)}+\frac{M^2\beta\left(\beta^2+3\beta+3\right)}{12T^2}\notag\\
&=\frac{M^2}{12T^2}+\frac{M^2\beta^2\left(\log^2(2)-1\right)^2}{2T^2\log^2(2)}.
\end{alignat}

\end{figure*}

For the general DDOP, a rectangular TD window given by $\bar{b}(t)=\Pi_{(N+2D)T}\left(t-\frac{(N+2D-1)T}{2}\right)$ acts as the envelope function of its TD representation, and the frequency response of $\bar{a}(t)$  acts as the envelope function of its FD representation. 

For the DDOP with the BTRRC sub-pulse, a rectangular TD window given by $b(t)=\Pi_{NT}\left(t-\frac{(N-1)T}{2}\right)$ acts as the envelope function of its TD representation, and the frequency response of BTRRC pulse acts as the envelope function of its FD representation. We note that when the sub-pulse in the DDOP is changed from the RRC pulse to the BTRRC pulse, the envelope function of its TD representation remains unchanged, but the envelope function of its FD representation changes. Based on this change, $\Delta F$ of the BTRRC pulse can be derived as \eqref{DeltaF_BTRRC}, which is given at the end of this page.

\begin{figure*}[t]
\centering\subfloat[\label{2a}$\Delta A$ \textrm{versus} $\beta$]{ \includegraphics[width=0.875\columnwidth]{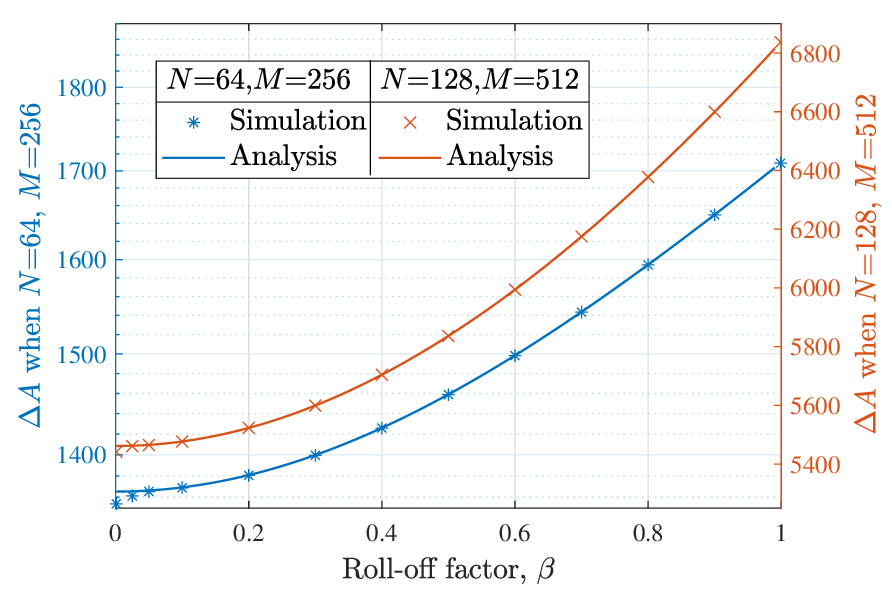}}
\subfloat[\label{2b}$\Delta T$ \textrm{versus} $\beta$]{ \includegraphics[width=0.875\columnwidth]{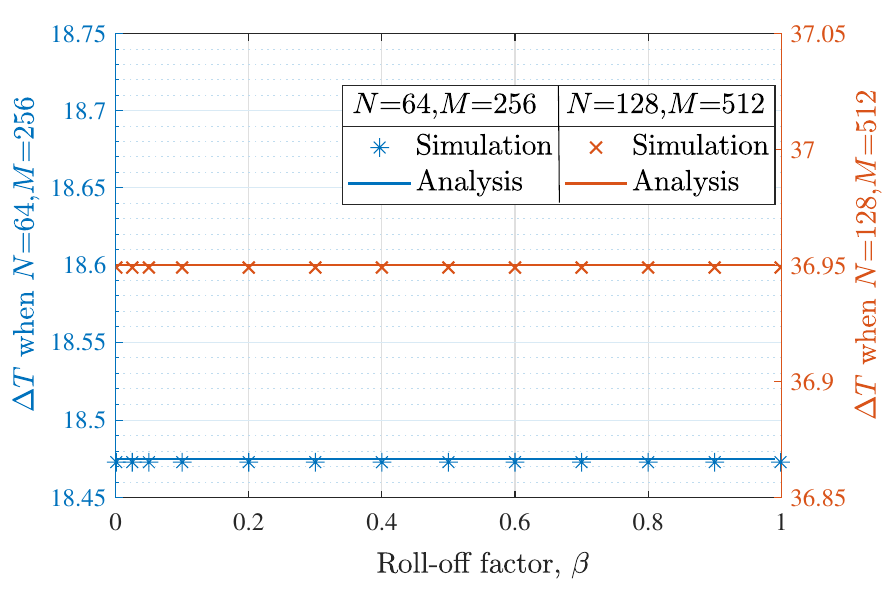}}
\hspace{0 mm} \vspace{-2mm}\subfloat[\label{2c}$\Delta F$ \textrm{versus} $\beta$]{\includegraphics[width=0.875\columnwidth]{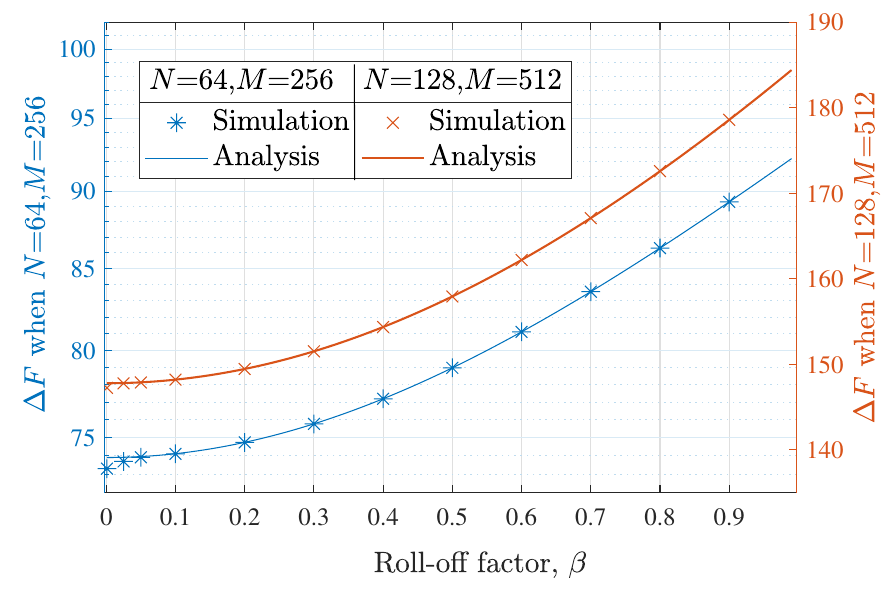}}
\subfloat[\label{2d}$\kappa$ \textrm{versus} $\beta$]{ \includegraphics[width=0.875\columnwidth]{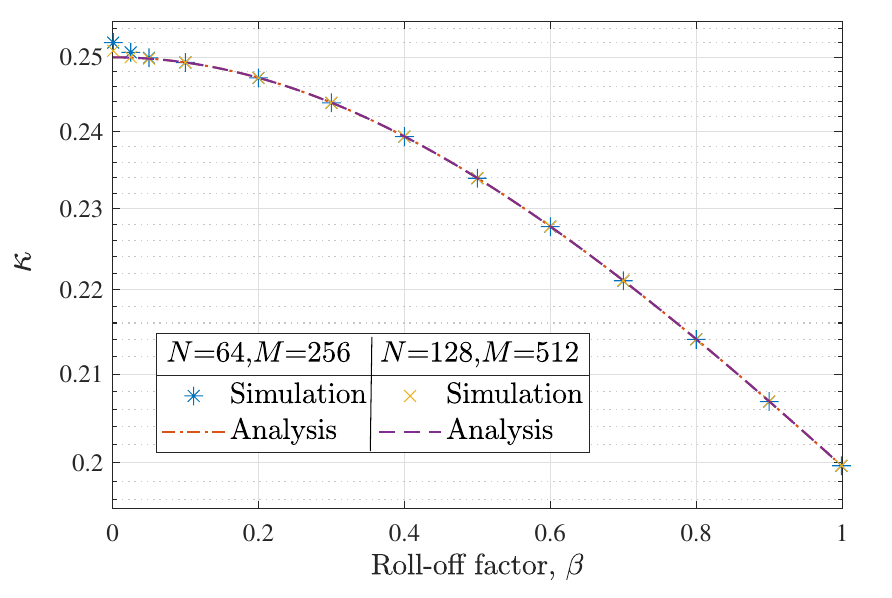}}
\vspace{1mm}
\caption{The TF localization metrics of the DDOP considered in \eqref{ut} versus the roll-off factor, $\beta$.}\label{Fig:Deltas_ut_Compare}
\end{figure*}

Using the aforementioned results and leveraging the insights obtained in \eqref{Equ:DeltaT_relF} and \eqref{Equ:DeltaF_relF} in Section \ref{Sec:WindowandFilter}, we deduce that $\Delta T$ and $\Delta F$ of the DDOP with BTRRC sub-pulse as
\begin{align}\label{DeltaT_DDOP_BTRRC}
\Delta T_{\textrm{DDOP}}\approx\frac{NT}{\sqrt{12}}
\end{align}
and
\begin{align}\label{DeltaF_DDOP_BTRRC}
\Delta F_{\textrm{DDOP}}\approx\frac{M}{T}\sqrt{\frac{1}{12}+\frac{\beta^2\left(\log^2(2)-1\right)^2}{2\log^2(2)}},
\end{align}
respectively. Finally, based on \eqref{DeltaT_DDOP_BTRRC} and \eqref{DeltaF_DDOP_BTRRC}, we obtain $\Delta A$ and $\kappa$ of the DDOP with BTRRC sub-pulse as
\begin{align}\label{DeltaA_DDOP_BTRRC}
\Delta A_{\textrm{DDOP}}\approx\frac{MN}{12}\sqrt{1
+\frac{6\beta^2\left(\log^2(2)-1\right)^2}{\log^2(2)}}
\end{align}
and
\begin{align}\label{k_DDOP_BTRRC}
\kappa_{\textrm{DDOP}}\approx\frac{NT^2}{M}\sqrt{\frac{\log^2(2)}
{\log^2(2)+6\beta^2\left(\log^2(2)-1\right)^2}},
\end{align}
respectively.

\color{black}

\section{TF Localization Metrics of Effective Pulses in OTFS} \label{Sec:OTFS}

\color{black}

In this section, we determine TF localization metrics
of effective pulses in OTFS. To this end, we leverage the insights obtained in Section \ref{Sec:WindowandFilter}. 

The OTFS modulation uses a two-step approach to generate the transmit signal~\cite{2017_WCNC_OTFS_Haddani}. In particular, the inverse symplectic finite Fourier transform is first applied on $X_{\mathrm{DD}}[n,m]$ to obtain the TF domain symbols 
\begin{align}\label{xt_OTFS1}
X_{\textrm{TF}}[\dot{n},l] &= \frac{1}{\sqrt{NM}}\sum_{n=0}^{N-1} \sum_{m=0}^{M-1} X_{\textrm{DD}}[n,m]e^{j2\pi\left(\frac{n\dot{n}}{N}-\frac{ml}{M}\right)},
\end{align}
where $\dot{n}\in\{0,1, \cdots,N-1\}$ and $l\in\{0,1, \cdots,M-1\}$. 
Next, a multicarrier modulator on the TF plane is used to generate the baseband OTFS signal from $X_{\textrm{TF}}[\dot{n},l]$ as
\begin{align}\label{xt_OTFS2}
x_{\mathrm{OTFS}}(t)&=\sum _{\dot{n}=0}^{N-1}\sum _{l=0}^{M-1}X_{\textrm{TF}}[\dot{n},l]e^{j2\pi l \Delta {f} (t-\dot{n}T)}g_{\textrm{TF}}(t-\dot{n}T),
\end{align}
where  $g_{\textrm{TF}}(t)$ is the transmit pulse of the multicarrier modulator on the TF plane.

It has to be noted that the ideal pulse, which causes neither ICI nor ISI, was proposed as $g_{\textrm{TF}}(t)$ in the original design of OTFS~\cite{2017_WCNC_OTFS_Haddani}. However, it cannot be realized in practice, thus making it impossible to determine the TF localization metrics of OTFS with an ideal pulse. 

As an alternative to the ideal pulse, the rectangular pulse $\frac{1}{\sqrt{T}}\Pi_T(t-\frac{T}{2})$ was considered as $g_{\textrm{TF}}(t)$ in~\cite{2018_TWC_Viterbo_OTFS_InterferenceCancellation}. Under such consideration, \eqref{xt_OTFS2} can be expressed in terms of $X_{\textrm{DD}}[n,m]$ as
\begin{align}\label{xt_OTFS3}
x_{\mathrm{OTFS}}(t)&=\sum_{n=0}^{N-1} \sum_{m=0}^{M-1} X_{\textrm{DD}}[n,m]\phi_{m,n}^{\textrm{OTFS}}(t),
\end{align}
where $\phi_{m,n}^{\textrm{OTFS}}(t)$ is the basis function of OTFS, which can be simplified as
\begin{align}\label{xt_OTFS4}
&\phi_{m,n}^{\textrm{OTFS}}(t)\notag\\
&=\frac{1}{\sqrt{NM}}\sum _{\dot{n}=0}^{N-1}e^{j2\pi \frac{n\dot{n}}{N}}b_{\mathrm{OTFS}}\left(t-\dot{n}T-m\frac{T}{M}\right)g_{\textrm{TF}}(t-\dot{n}T),
\end{align}
where 
\begin{align}\label{xt_OTFS5}
&b_{\mathrm{OTFS}}(t)=e^{j(M-1) \frac{\pi}{T}t}\frac{\mathrm{sin}\left(M\frac{\pi}{T}t\right)}{\mathrm{sin}\left(\frac{\pi}{T}t\right)}.
\end{align}
We clarify that although $g_{\textrm{TF}}(t)=\frac{1}{\sqrt{T}}\Pi_T(t-\frac{T}{2})$ is directly used as the pulse in the OTFS representation/implementation, it is the basis function of OTFS $\phi_{m,n}^{\textrm{OTFS}}(t)$ in \eqref{xt_OTFS4} that effectively carries the $(m,n)$-th symbol in $X_{\mathrm{DD}}[m,n]$. Thus, we characterize the TF localization metrics of pulses of OTFS using the basis functions of OTFS $\phi_{m,n}^{\textrm{OTFS}}(t)$.

For $\phi_{m,n}^{\textrm{OTFS}}(t)$ in \eqref{xt_OTFS4}, a rectangular TD window $\Pi_{T}\left(t-\frac{NT}{2}-m\frac{T}{M}\right)$ acts as the envelope function of its TD representation. Additionally, the frequency response of $\phi_{m,n}^{\textrm{OTFS}}(t)$, can be characterized as a train of tones with an envelope function, where the envelope is a rectangular $\Pi_{\frac{M}{T}}\left(f\right)$ with some out-of-band emission (OOBE) depending on $m$. Based on these envelope functions 
and leveraging the insights obtained from Section \ref{Sec:WindowandFilter}, we deduce $\Delta T$ and $\Delta F$ of the basis functions of OTFS $\phi_{m,n}^{\textrm{OTFS}}(t)$ as
\begin{align}\label{DeltaT_OTFS}
\Delta T_{\textrm{OTFS}} &\approx\frac{NT}{\sqrt{12}}
\end{align}
and
\begin{align}\label{DeltaF_OTFS}
\Delta F_{\textrm{OTFS}}\approx\frac{M}{T\sqrt{12}},
\end{align}
respectively. Finally, based on \eqref{DeltaT_OTFS} and \eqref{DeltaF_OTFS}, we obtain $\Delta A$ and $\kappa$ of the basis functions of OTFS $\phi_{m,n}^{\textrm{OTFS}}(t)$ as
\begin{align}\label{DeltaA_OTFS}
\Delta A_{\textrm{OTFS}}\approx\frac{MN}{12}
\end{align}
and
\begin{align}\label{k_OTFS}
\kappa_{\textrm{OTFS}}\approx\frac{NT^2}{M},
\end{align}
respectively.
\color{black}

\color{black}
By comparing \eqref{DeltaT_DDOP} with \eqref{DeltaT_OTFS},  it is evident that the DDOP $u(t)$ and effective pulses in OTFS $\phi_{m,n}^{\textrm{OTFS}}(t)$ share the same $\Delta T$. This is because the envelope functions of their TD representations are the same. However, by comparing \eqref{DeltaF_DDOP} with \eqref{DeltaF_OTFS}, it can be observed that their $\Delta F$ values differ due to the distinct envelope functions of their FD representations.
\color{black}

\color{black}
Moreover, we note that the basis functions of OTFS $\phi_{m,n}^{\textrm{OTFS}}(t)$ can experience high OOBE  when $m=0$ or $m=M-1$, $\forall n\in\{0,1,\cdots,N-1\}$. Given this and the fact that the results in \eqref{DeltaF_OTFS}-\eqref{k_OTFS} are derived by ignoring the impact of OOBE,  the accuracy of the results for OTFS in \eqref{DeltaF_OTFS}-\eqref{k_OTFS} may be compromised when $m=0$ or $m=M-1$, $\forall n\in\{0,1,\cdots,N-1\}$.
\color{black}

\section{Numerical Results}\label{Sec:Num}

In this section, we present numerical results to evaluate our analytical results. Unless specified otherwise, we consider $M=256$, $N=64$, a normalized $T$, i.e., $T=1$, the roll-off factor $\beta=0.1$, and let $Q\approx 0.05 M$ to ensure \textcolor{black}{$T_a\approx 0.1 T$ such} that the duration condition $T_a\ll T$ is met for the considered DDOP in \eqref{ut}~\cite{2022_ICC_JH_ODDM,2022_TWC_JH_ODDM,2022_GC_JH_ODDMPulse,2023_ODDM_TCOM}. 
\color{black}
We clarify that although analytical results in previous sections are derived while ignoring the side-lobes of \(A(f)\), numerical results in this section are determined while considering all the side-lobes of \(A(f)\) within the spectrum \(-5\frac{M}{T}\) to \(5\frac{M}{T}\).
\color{black}


\begin{figure}[t]
\centering
\includegraphics[width=0.875\columnwidth]{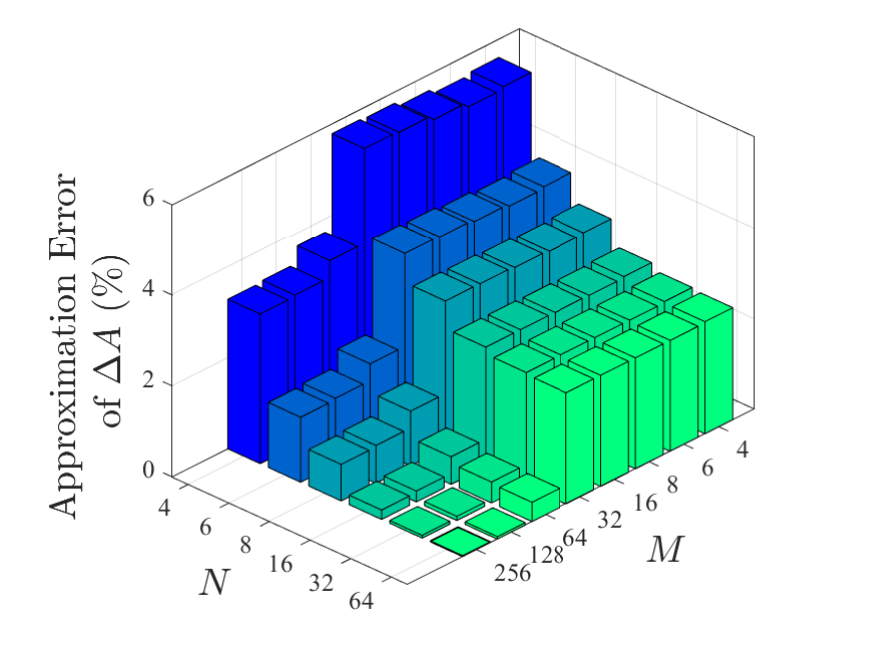}
\caption{\textcolor{black}{Illustration of the approximation error of $\Delta A$ for different values of $M$ and $N$.}}\label{FigF:ApproxError} 
\end{figure}

\begin{figure*}[t]
\centering\subfloat[\label{3a}$\Delta A$]{ \includegraphics[width=0.875\columnwidth]{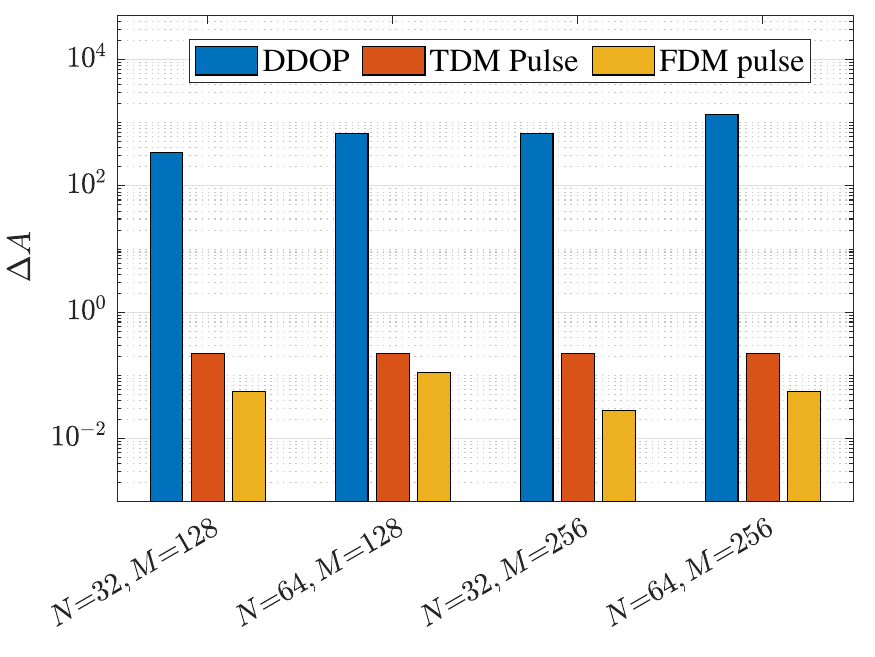}}
\subfloat[\label{3b}$\Delta T$]{ \includegraphics[width=0.875\columnwidth]{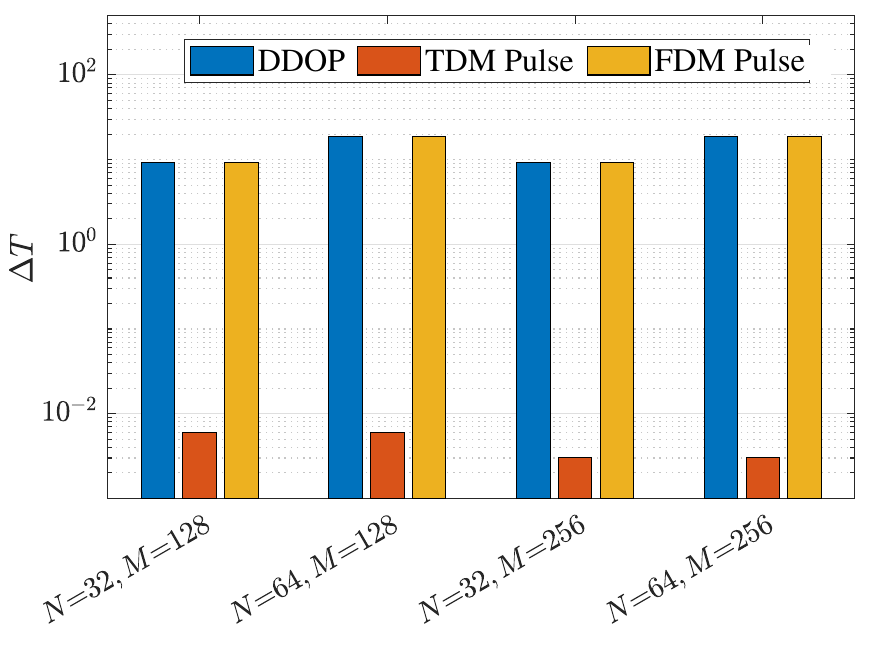}}
\vspace{-2mm} \subfloat[\label{3c}$\Delta F$]{\includegraphics[width=0.875\columnwidth]{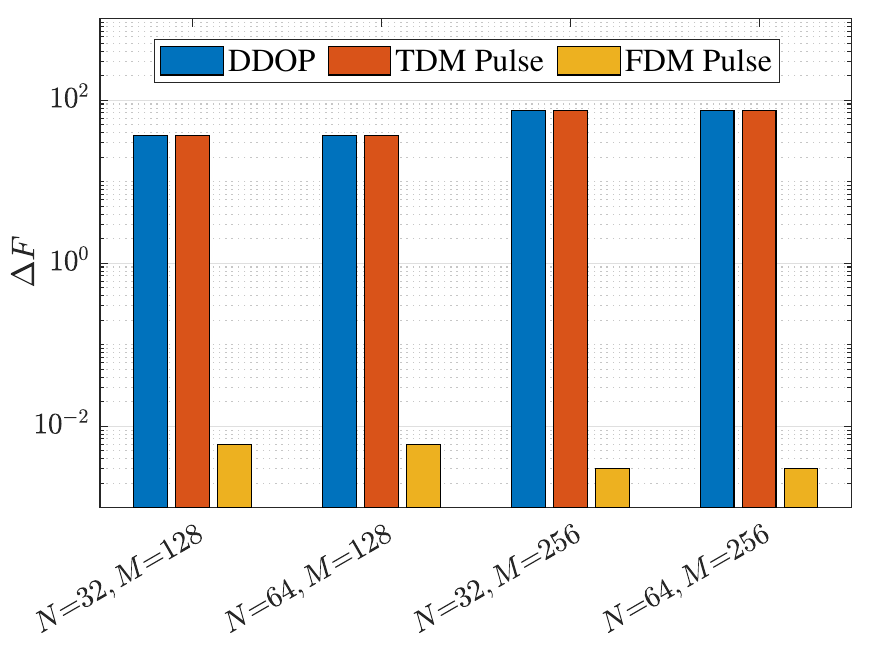}}
\subfloat[\label{3d}$\kappa$]{ \includegraphics[width=0.875\columnwidth]{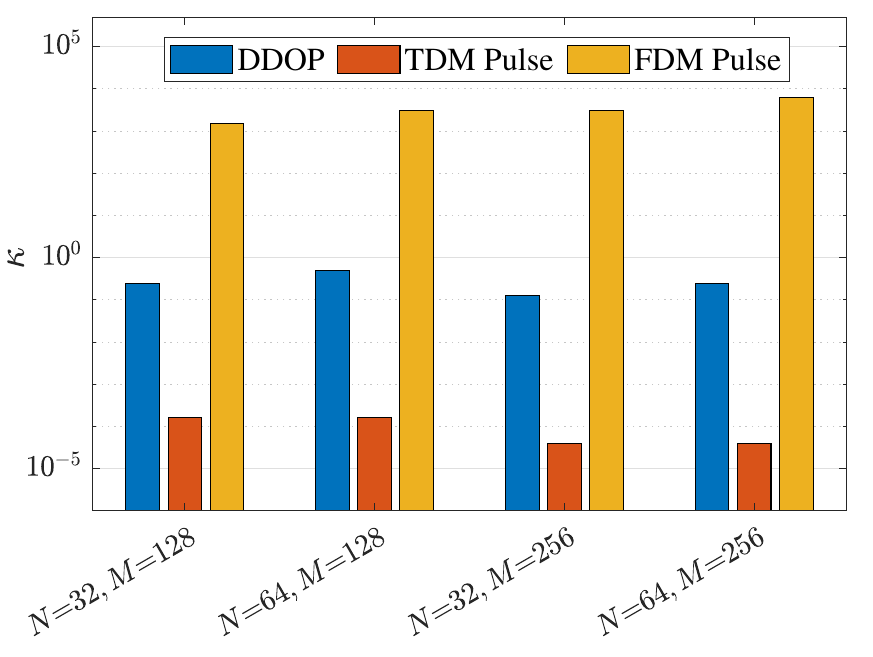}}
\caption{A comparison of the TF localization metrics of the DDOP with those of the pulses used in TDM and FDM schemes.}\label{Fig:DDOPTDMFDM_Compare}
\end{figure*}

In Fig. \ref{Fig:Deltas_ut_Compare}, we verify the derived results in \emph{Theorem \ref{Thr:IORCPOTFS-VCP-ECU}}. To this end, we plot (i) the TF localization metrics presented in \emph{Theorem \ref{Thr:IORCPOTFS-VCP-ECU}} and (ii) the numerically calculated TF localization metrics of the simulated DDOP, versus $\beta$ for different values of $M$ and $N$. We observe that for all values of $M$, $N$, and $\beta$, the analytical results for TF localization metrics match well with the simulation results. For the considered parameters, the maximum percentage difference between the analytical and simulation results for $\Delta A$, $\Delta T$, $\Delta F$, and $\kappa$ are $0.8355\%$, $0.0122\%$, $0.823\%$, and $0.818\%$, respectively. This observation shows the correctness of our derived results in \emph{Theorem \ref{Thr:IORCPOTFS-VCP-ECU}}. We further clarify that when deriving $\Delta F$, we utilize the frequency response of the RRC pulse as the frequency response of the truncated RRC pulse while ignoring the impact of truncation. This approximation leads to the marginal difference (e.g., less than $1\%$) between the analytical and simulation results for $\Delta A$, $\Delta F$, and $\kappa$ when $Q$ and/or $\beta$ is low. In addition, we note that these marginal differences vanish as $Q$ and/or $\beta$ increases, which further validates our derived results in \emph{Theorem \ref{Thr:IORCPOTFS-VCP-ECU}}.
\textcolor{black}{Finally, we observe in Figs.~\ref{Fig:Deltas_ut_Compare} (b) and (c) that when  $\beta$ changes, $\Delta T$ remains unchanged while  $\Delta F$ changes. This behavior can be explained by the derived relationships between $\Delta T$ and $\Delta F$ of the DDOP and those of the rectangular TD window $b(t)$ and the sub-pulse $a(t)$ used for its generation, given in Section~\ref{Sec:GDDOP}. In particular, when $\beta$ changes, the characteristics of $b(t)$ remain unchanged, but those of $a(t)$ change. Due to this and the fact that $\Delta T$ of $b(t)$ dictates $\Delta T$ of the DDOP and $\Delta F$ of $a(t)$ dictates $\Delta F$ of the DDOP, $\Delta T$ remains unchanged while  $\Delta F$ changes for varying $\beta$.}

\textcolor{black}{In Fig. \ref{FigF:ApproxError}, we explore the validity of our derived results in \emph{Theorem \ref{Thr:IORCPOTFS-VCP-ECU}}. To this end,   we plot the approximation error of $\Delta A$ for different values of $M$ and $N$. We observe 
that the approximation error of $\Delta A$ is extremely low for high and moderate values of $M$ and $N$, i.e., the error is less than $1\%$ for $M\geq 64$ and  $N\geq 16$. However, as $M$ and/or $N$ decreases, the approximation error of $\Delta A$ increases, reaching a maximum of $6\%$ for extremely small values of $M$ and $N$, such as $M=N=4$. Despite this observation, it is important to acknowledge that in order to explore time and frequency diversities, practical ODDM systems would typically employ large $M$ and $N$, e.g., $M=256$ and $N=64$ \cite{2022_TWC_JH_ODDM}. In such scenarios, the approximation error of $\Delta A$  is extremely small, demonstrating the significance of our derived results.}


In Fig. \ref{Fig:DDOPTDMFDM_Compare}, we compare the numerically calculated TF localization metrics of the simulated DDOP, with those of the pulses used in TDM and FDM schemes for different values of $M$ and $N$. Based on this comparison, we find that the conclusions drawn in Section \ref{Sec:Dis} regarding the energy spread of the DDOP in comparison to the pulses used in TDM and FDM schemes are correct.
Particularly, based on the values of $\Delta A$ in Fig. \ref{Fig:DDOPTDMFDM_Compare}(a), we verify that the energy of the DDOP in the joint TF domain is more widely spread compared to those of the pulses used in TDM and FDM schemes. Second, based on the values of $\Delta T$ in Fig. \ref{Fig:DDOPTDMFDM_Compare}(b), we verify that the energy of the DDOP in the TD is more widely spread than that of the pulse used in the TDM scheme. Third, based on the values of $\Delta F$ in Fig. \ref{Fig:DDOPTDMFDM_Compare}(c), we verify that the energy of the DDOP in the FD is more widely spread than that of the pulse used in the FDM scheme. Finally, based on the values of $\kappa$ in Fig. \ref{Fig:DDOPTDMFDM_Compare}(d), we verify that the amount of stretch of the DDOP along the TD relative to the FD is higher than that of the pulse used in the TDM scheme but lower than that of the pulse used in the FDM scheme. In addition, we observe from Fig. \ref{Fig:DDOPTDMFDM_Compare}(d) that unlike the pulses used in TDM and FDM schemes which always exhibit a wider energy spread in one dimension (TD or FD) compared to the other, DDOP provides a good tradeoff between its energy spread in the TD and FD.

\begin{figure*}[t]
\centering\subfloat[\label{4a}$\Delta A$ \textcolor{black}{for general DDOP}]{ \includegraphics[width=0.875\columnwidth]{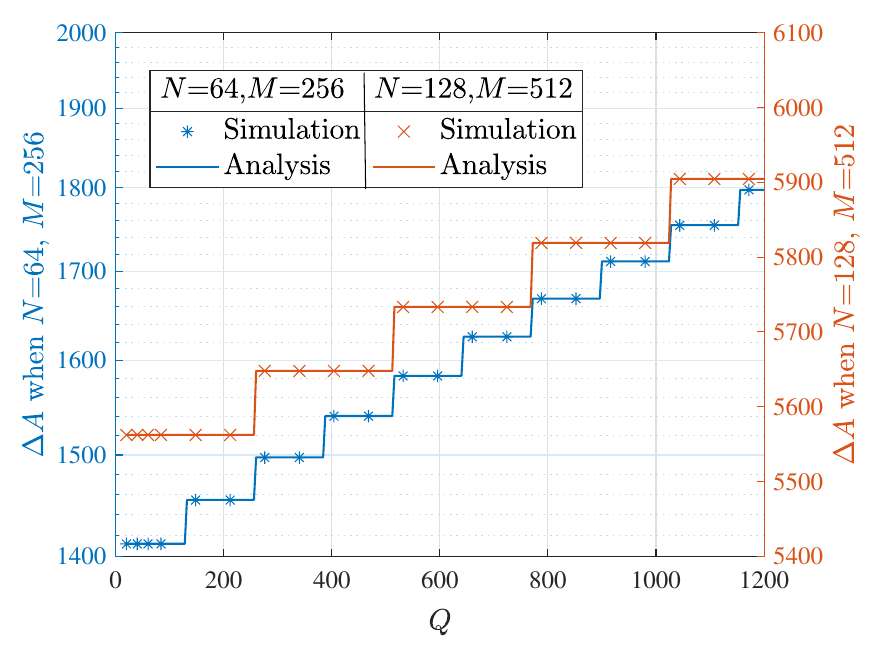}}
\subfloat[\label{4b}$\Delta T$ \textcolor{black}{for general DDOP}]{ \includegraphics[width=0.875\columnwidth]{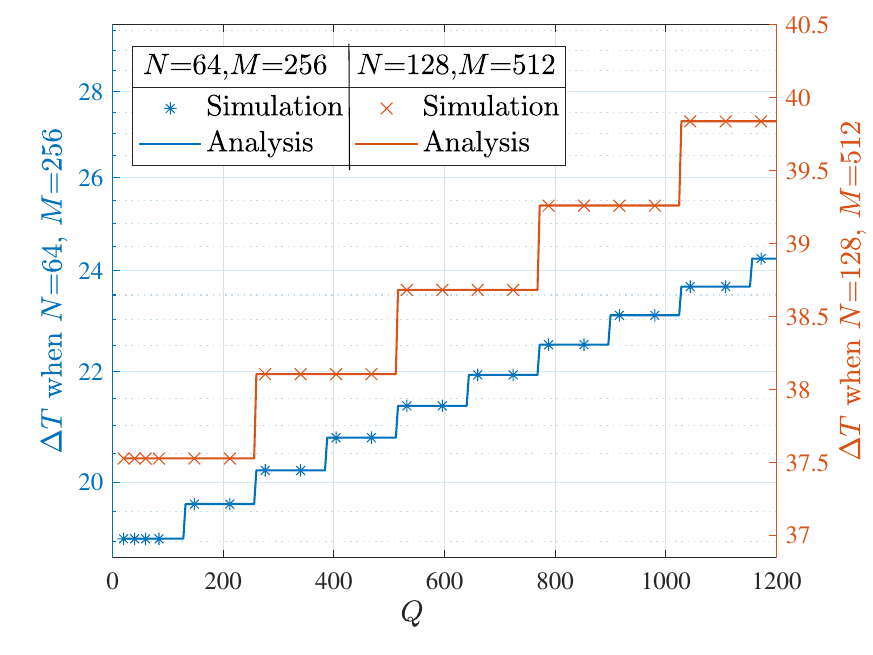}}
\vspace{-2mm} \subfloat[\label{4c}$\Delta F$ \textcolor{black}{for general DDOP}]{\includegraphics[width=0.875\columnwidth]{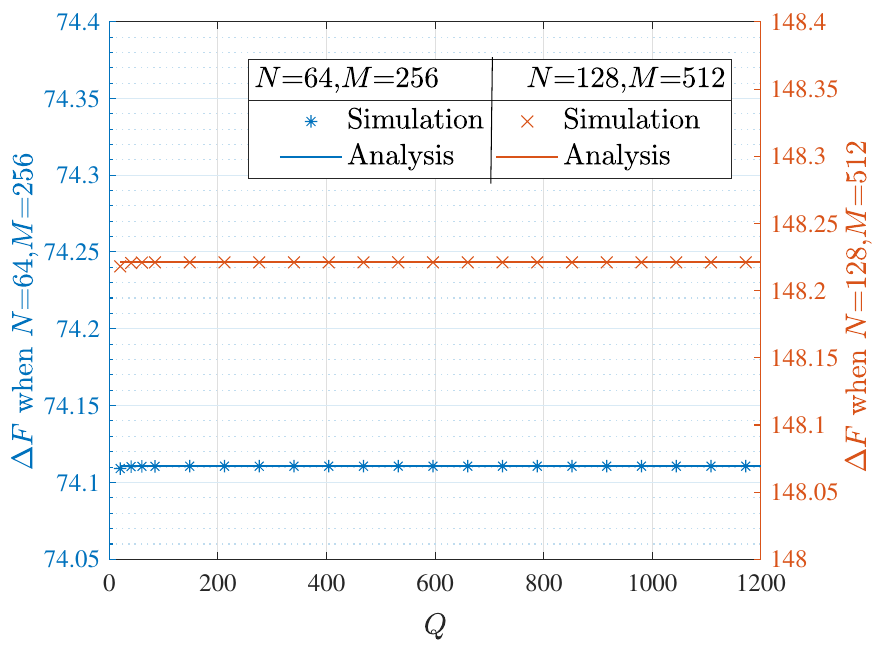}}
\subfloat[\label{4d}$\kappa$ \textcolor{black}{for general DDOP}]{ \includegraphics[width=0.875\columnwidth]{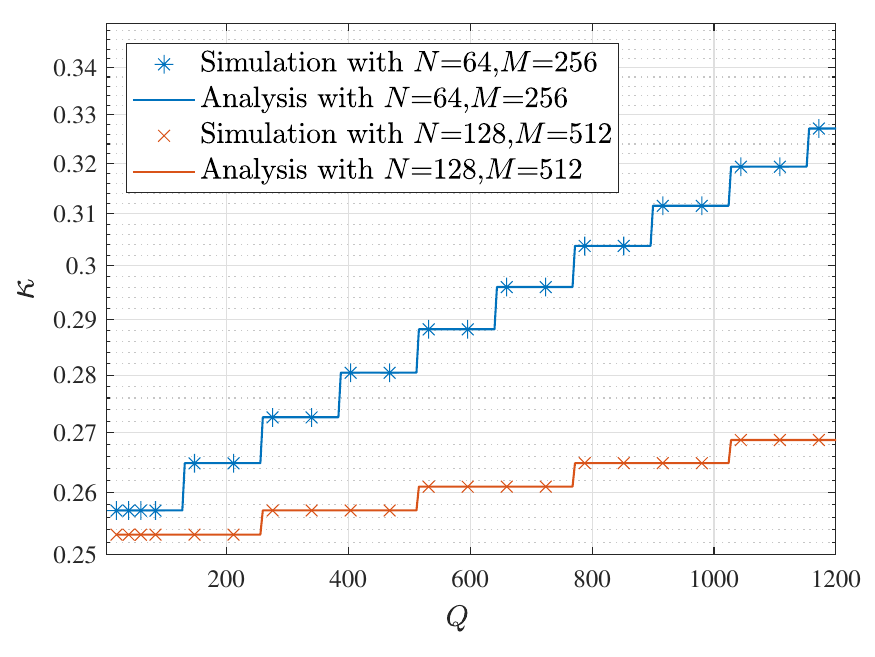}}
\caption{The TF localization metrics of the generalized design of the DDOP given in \eqref{ut_general} versus $Q$.}\label{Fig:Deltas_utgeneral_Compare}
\end{figure*}

\begin{figure*}[t]
\centering\subfloat[\label{5a}$\Delta T$ \textrm{versus} $\beta$]{ \includegraphics[width=0.875\columnwidth]{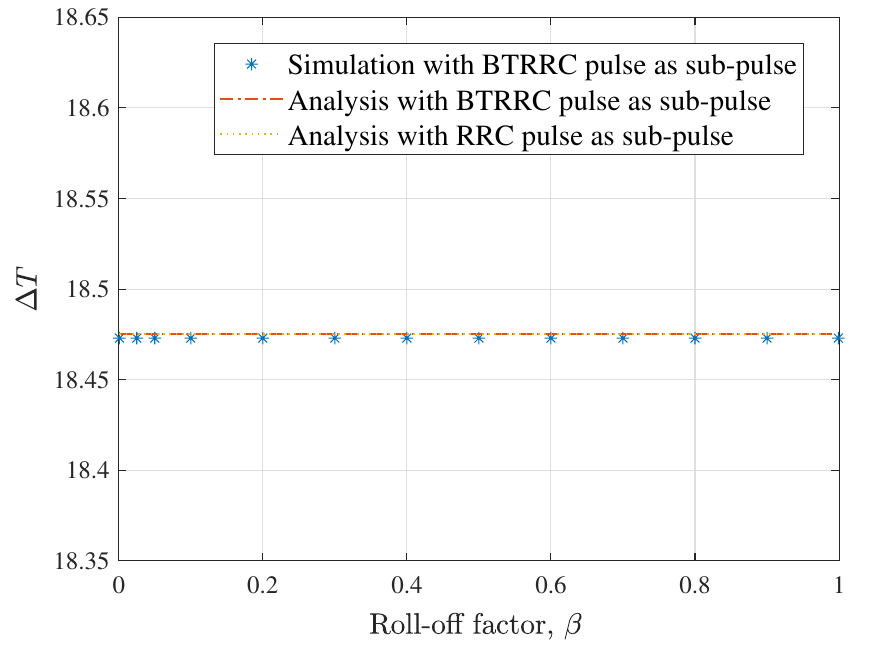}}
\subfloat[\label{5b}$\Delta F$ \textrm{versus} $\beta$]{ \includegraphics[width=0.875\columnwidth]{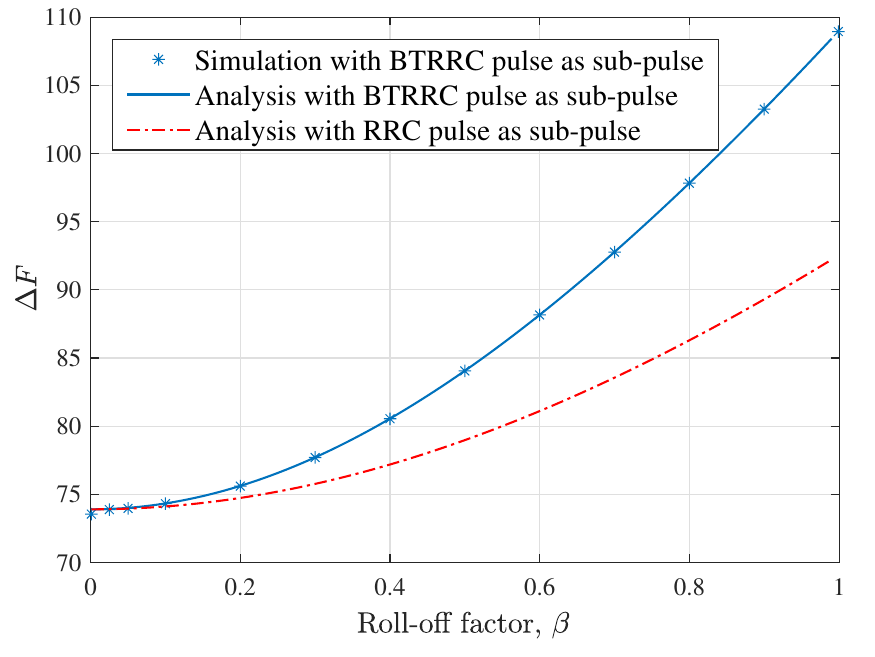}}
\vspace{-2mm} \subfloat[\label{5c}$\Delta A$ \textrm{versus} $\beta$]{\includegraphics[width=0.875\columnwidth]{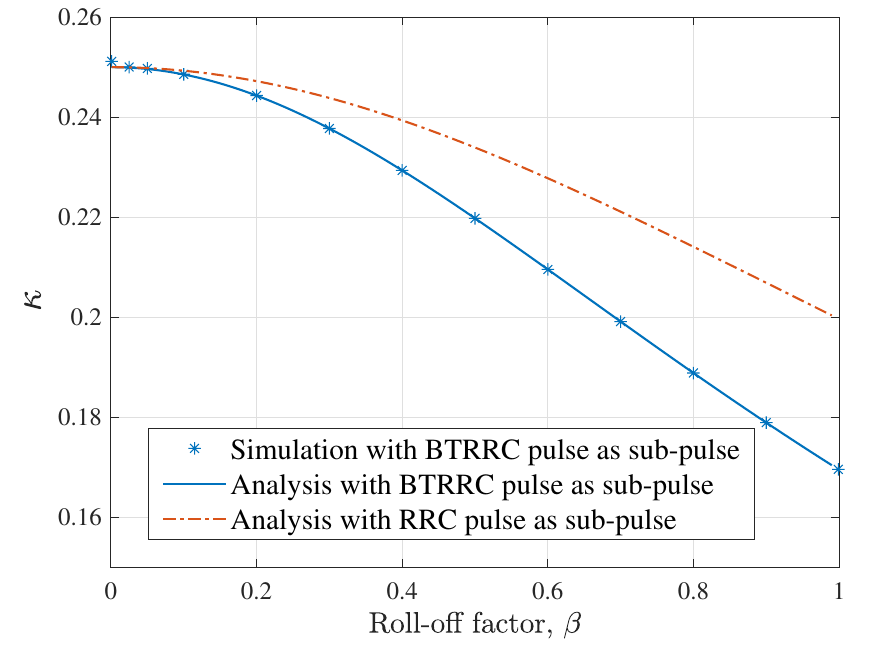}}
\subfloat[\label{5d}$\kappa$ \textrm{versus} $\beta$]{ \includegraphics[width=0.875\columnwidth]{FigDd_V1-eps-converted-to.pdf}}
\caption{The TF localization metrics of the DDOP with different sub-pulses versus the roll-off factor, $\beta$.}\label{Fig:DDOPwithBTSRRC}
\end{figure*}

Next, in Fig. \ref{Fig:Deltas_utgeneral_Compare},  we verify the derived results for the generalized design of the DDOP given in Section \ref{Sec:GDDOP}. To this end, we plot (i) the TF localization metrics presented in \eqref{DeltaT_GDDOP}--\eqref{k_GDDOP} in Section \ref{Sec:GDDOP} and (ii) the numerically calculated TF localization metrics of the simulated general DDOP, versus $Q$ for different values of $M$ and $N$. We observe that for all values of $M$, $N$, and $Q$, the analytical results for TF localization metrics match well with those from simulations. This shows the correctness of our analysis for the generalized design of the DDOP in Section \ref{Sec:GDDOP}. We also observe a step-wise increase in $\Delta A$, $\Delta T$, and $\kappa$ in Figs. \ref{Fig:Deltas_utgeneral_Compare}(a), (b), and (d), respectively, as $Q$ increases. These are due to the step-wise increase in the lengths of the CP and CS, which occurs in the general DDOP, as a result of increasing $Q$, since $D=\lceil\frac{T_a}{T}\rceil=\lceil\frac{2Q}{M}\rceil$.

Finally, in Fig. \ref{Fig:DDOPwithBTSRRC}, we evaluate the results derived in Section \ref{Sec:DDOP_BTRRC} for the DDOP with the BTRRC sub-pulse. To this end, we plot the analytical and simulation values of the TF localization metrics of the DDOP with the BTRRC sub-pulse versus $\beta$. For comparison, the analytical values of the TF localization metrics of the DDOP with the RRC sub-pulse are also plotted. We first observe that for all values of $Q$, the analytical results for the TF localization metrics match well with those from simulations. This shows the correctness of our derived results for the DDOP with the BTRRC sub-pulse. We also observe that although $\Delta T$ of the DDOP with the RRC sub-pulse is comparable to that with the BTRRC pulse, $\Delta A$ and $\Delta F$ of the DDOP with the RRC sub-pulse are lower than those with the BTRRC sub-pulse. This observation shows that the DDOP can achieve a lower energy spread in the joint TF domain and the FD, if the RRC pulse instead of the BTRRC pulse is utilized in its design, especially when the roll-off factor of the sub-pulse $\beta$ is high.

\section{Conclusions}\label{Sec:Concl}

In this work, we studied the TF localization characteristics of the prototype pulse of ODDM modulation, which is known as the DDOP. We first derived the TF localization metrics, namely, the TFA, as well as the time dispersion, frequency dispersion, and direction parameter of the DDOP. Based on the derived results, we provided insights into the energy spread of the DDOP in the joint TF domain, TD, and FD. 
Thereafter, we discussed the potentials offered by the DDOP due to its energy spread. Particularly, we highlighted that harnessing both time and frequency diversities necessitates the wide energy spread of the DDOP. 
We also pointed out that the wide energy spread in the TD and FD makes DDOP well-suited for sensing applications which require fine resolutions. We further presented the relationship between the time and frequency dispersions of the DDOP and those of the envelope functions of DDOP’s TD and FD
representations.
In addition, we determined the TF localization metrics of the recently proposed generalized design of the DDOP. Aided by numerical and simulation results, we corroborated our analysis and showed that the DDOP can achieve a lower energy spread in the joint TF domain and FD, if the RRC pulse, rather than the BTRRC pulse, is utilized as its sub-pulse.

\appendices

\section{Lemma in Section \ref{Sec:DeltaTFofDDOPDerivation}}\label{App:Lemmas}

\begin{lemma}\label{Lem:t2shiftAll}
Considering an even function $\mathcal{X}:\rho\rightarrow\mathcal{X}(\rho)$, where $\rho\in \mathbb{R}$ and $\mathcal{X}(\rho)\in \mathbb{C}$, we find that
\begin{align}\label{t2shift}
\!\!\int_{-\infty}^{\infty} \!\!\!\!\!\!\rho^2|\mathcal{X}(\rho_{\alpha} \rho-\rho_{\gamma})|^2 \mathrm{d}\rho =\!\int_{-\infty}^{\infty} \!\!\!\!\!\! \rho^2|\mathcal{X}(\rho_{\alpha}\rho)|^2 \mathrm{d}\rho+\frac{\rho^2_{\gamma}}{\rho^2_{\alpha}} \mathcal{E}_{\mathcal{X}},
\end{align}
where 
$\mathcal{E}_{\mathcal{X}}=\int_{-\infty}^{\infty}|\mathcal{X}(\dot{\rho})|^2 \mathrm{d}\dot{\rho}$ and $\rho_{\alpha},\rho_{\gamma}\in \mathbb{R}$.
\end{lemma}


\end{document}